\documentclass{article}

\usepackage{hyperref} 

\usepackage{arxiv}
\usepackage{float} 

\usepackage[utf8]{inputenc}
\usepackage[T1]{fontenc}

\hypersetup{
    linktoc=all,
    colorlinks=true,
    linkcolor=black,
    citecolor=black,
    urlcolor=black,
}

\usepackage{natbib}
\let\cite\citep

\usepackage{svg}
\usepackage{amsthm}
\usepackage{amsmath}
\usepackage{amssymb}
\newtheorem{remark}{Remark}

\usepackage{bbm}
\usepackage{wasysym}
\usepackage{mathtools}
\usepackage{enumitem}

\usepackage{lineno}

\usepackage{graphicx}
\graphicspath{{figure/}} %Setting the graphicspath

\usepackage{subcaption}

\usepackage{multirow}

\usepackage{xpatch}
\xpretocmd{\eqref}{Eq.~}{}{}

\DeclareUnicodeCharacter{2212}{-}

\DeclareMathOperator*{\argmin}{arg\,min}

\newcommand{\romnum}[1]{\uppercase\expandafter{\romannumeral #1\relax}}

\title{Radius selection using kernel density estimation for the computation of nonlinear measures} 

\date{}
\author{\href{https://orcid.org/0000-0002-7558-2071}{\includegraphics[scale=0.06]{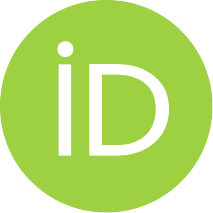}\hspace{1mm}Johan Medrano}      \vspace{2mm}   \\
	CNRS - University of Montpellier LIRMM, UMR5506 \\
	Interactive Digital Human \\
        Montpellier, France \vspace{2mm}  \\ 
	\texttt{\href{email:johan.medrano@ucl.ac.uk}{johan.medrano@ucl.ac.uk}} \\
        \And 
        \href{https://orcid.org/0000-0001-9033-9742}{\includegraphics[scale=0.06]{orcid.pdf}\hspace{1mm}Abderrahmane Kheddar} \vspace{2mm}  \\
	CNRS - AIST Joint Robotics Laboratory, IRL3218 \\
	Tsukuba 305-8560, Japan  \vspace{2mm}  \\
	CNRS - University of Montpellier LIRMM, UMR5506 \\
	Interactive Digital Human \\
        Montpellier, France  
        \And
        \href{https://orcid.org/0000-0002-6647-612X}{\includegraphics[scale=0.06]{orcid.pdf}\hspace{1mm}
        Annick Lesne} \vspace{2mm} \\ 
        Sorbonne Université, CNRS \\ 
        Laboratoire de Physique Théorique de la Matière Condensée, LPTMC \\
        F-75252 Paris, France 
        \And
        \href{https://orcid.org/0000-0001-8925-3546}{\includegraphics[scale=0.06]{orcid.pdf}\hspace{1mm}Sofiane Ramdani} \vspace{2mm} \\
	CNRS - University of Montpellier LIRMM, UMR5506 \\
	Interactive Digital Human \\
        Montpellier, France 
 }

\begin{document}

% \begin{frontmatter}

% \author{Johan Medrano}
% \email[]{johan.medrano@lirmm.fr}
% \affiliation{LIRMM, CNRS UMR 5506, University of Montpellier, F-34095 Montpellier, France
% }%

% \author{Abderrahmane Kheddar}%
%  \email[]{kheddar@lirmm.fr}
% \affiliation{LIRMM, CNRS UMR 5506, University of Montpellier, F-34095 Montpellier, France
% }
% \affiliation{CNRS-AIST Joint Robotics Laboratory, IRL, Tsukuba 305-8560, Japan}

% \author{Annick Lesne}
% \email[]{annick.lesne@sorbonne-universite.fr}
% \affiliation{Sorbonne Université, CNRS, Laboratoire de Physique Théorique de la Matière Condensée, LPTMC, F-75252 Paris, France}
% \affiliation{Institut de Génétique Moléculaire de Montpellier, University of Montpellier, CNRS, F-34293 Montpellier, France}

% \author{Sofiane Ramdani}
% \email[]{sofiane.ramdani@umontpellier.fr}
% \affiliation{LIRMM, CNRS UMR 5506, University of Montpellier, F-34095 Montpellier, France
% }

% \address[1]{LIRMM, CNRS UMR 5506, University of Montpellier, Montpellier, France}
% \address[2]{CNRS-AIST Joint Robotics Laboratory, IRL, Tsukuba, Japan}
% \address[3]{Sorbonne Université, CNRS, Laboratoire de Physique Théorique de la Matière Condensée, LPTMC,  F-75005 Paris, France}
% \address[4]{Institut de Génétique Moléculaire de Montpellier, University of Montpellier, CNRS, Montpellier, France}

% \cortext[cor1]{Corresponding author}

\maketitle

\begin{abstract}
    When nonlinear measures are estimated from sampled temporal signals with finite-length, a radius parameter must be carefully selected to avoid a poor estimation. These measures are generally derived from the correlation integral which quantifies the probability of finding neighbors, i.e.~pair
    of points spaced by less than the \emph{radius} parameter. While each nonlinear measure comes with several specific empirical rules to select a radius value, we provide a systematic selection method. We show that the optimal radius for nonlinear measures can be approximated by the optimal 
    \emph{bandwidth} of a Kernel Density Estimator (KDE) related to the correlation sum. The KDE framework provides non-parametric tools to approximate a density function from finite samples (e.g.\ histograms) and optimal methods to select a smoothing parameter, the bandwidth (e.g.\ bin width in histograms). We use results from KDE to derive a closed-form expression for the optimal radius. The latter is used to compute the correlation dimension and to construct recurrence plots yielding an estimate of Kolmogorov-Sinai entropy. We assess our method through numerical experiments on signals generated by nonlinear systems and experimental electroencephalographic time series.
\end{abstract}


% \begin{quotation}
%     Nonlinear measures computed from discrete time series are widely used 
%     to characterize and identify their underlying dynamics. Such measures include
%     entropies, dimensions and indices that are derived from recurrence plots. These indices, 
%     all based on the fundamental concept of correlation sum, have been shown to be 
%     effective in distinguishing  dynamical processes based on experimental data in various 
%     fields, e.g. mechanics, physiology, etc. One crucial parameter involved in their estimation
%      is the radius used to define neighbors in the state space. Although many rule of thumbs are available, 
%     there is a need of well-grounded theoretical criteria to correctly select this radius parameter. 
%     We propose to address this issue using Kernel Density
%     Estimation (KDE). We first demonstrate the theoretical link between the
%      correlation sum and the KDE framework. We then derive a loss function,
%      whose optimization is equivalent to the minimization of the
%     mean integrated squared error of a kernel density estimator, leading to a
%     closed-form criterion for the radius parameter selection. These
%     findings moreover show how the estimator bias-variance trade-off determines a range for the radius values.
%     Numerical experiments on both simulated
%     (chaotic, corrupted by additive noise) and real-world data are presented to
%     assess our approach. 
% \end{quotation}

\section{Introduction}
    %%% General intro
    Nonlinearity and chaos govern a wide variety of systems. They are found in neurons firing patterns~\cite{faure2001there} and related electrophysiological signals~\cite{freeman2003evidence}, and in unpredictable changes of Earth climate~\cite{ghil2008climate}, to cite few examples.  
    Nonlinear measures of such systems are made more accurate thanks to an increasing interest in numerical tools suitable for nonlinear phenomena. 
    Indeed, data generated by such systems are more suitable to nonlinear time series analysis, which provide complementary information to traditional linear methods such as power spectrum analysis \cite{yang2018recurrence}.    

    %%% neighbors/"complexity measures"
    Our work focuses on estimating various metrics, measures (in the sense of quantitative indices), and invariants that rely on the 
    computation of a correlation sum. The correlation sum is the estimator of the correlation integral, which is the mean probability that two points of the phase space trajectory of a dynamical system are neighbors \cite{ott2002chaos}, i.e.\ the mean probability that  
    their distance is less than a parameter called \emph{radius}, \emph{threshold} or \emph{tolerance} depending on the application domain. 
    The correlation sum captures important aspects of the nonlinear dynamics. Therefore, it is a fundamental quantity in various nonlinear measures: correlation dimension~\cite{grassberger1983characterization}, 
    Kolmogorov-Sinai entropy~\cite{grassberger1983estimation, eckmann1985ergodic, faure1998new}, its approximate versions 
    ApEn~\cite{pincus1991approximate} and SampEn~\cite{richman2000physiological}, R\'enyi's entropies \cite{principe2010information, singh2011information}, recurrence plots~\cite{eckmann1987recurrence}, \cite{marwan2007recurrence} and related metrics of 
    recurrence quantification analysis~\cite{grendar2013strong}, etc.
    
    In different nonlinear measures, the radius appears either as a variable or as a parameter. For 
    instance, the correlation dimension is computed by estimating a scaling factor on a logarithmic plot of
    the correlation sum versus the radius. In contrast, a recurrence plot 
    displays neighboring points on a black and white image and 
    requires to fix the radius parameter beforehand. In both cases, the radius is selected as small as possible. As a correlation sum computed from a finite-length time series will likely tend to 0 together with the radius parameter,
    the challenge is to identify a radius range corresponding to a statistically useful distribution of neighbors. Eckmann and Ruelle~\cite{eckmann1985ergodic} (Section \romnum{5}.A.1.a.) refer to it as a ``meaningful range'' for the radius parameter. 
    In our approach, we first derive an expression of the optimal radius. Then, we introduce a range to select a radius parameter or to study the properties of a function of the radius.
    
    %%% neighbors -> kde 
   Several empirical rules exist to select a value or a range of values for the radius; however, they generally focused on a particular nonlinear 
   measure~\cite{webber2015recurrence, zbilut1992embeddings, pincus1991approximate}. 
    Here, we introduce a method which can be applied to any nonlinear measure derived from the correlation sum.
    Observing that log-correlation sums are particularly used in nonlinear indices and 
    that relative error arises from logarithmic error terms, we focus on minimizing a relative error between the correlation sum and the correlation integral. We show that minimizing the relative error term is equivalent to minimizing a well-known error used in the framework of Kernel Density Estimation (KDE), widely studied in statistics~\cite{silverman1986density} and in signal processing~\cite{gunduz2009correntropy, singh2011information}.
    
    KDE denotes a family of non-parametric density estimation methods which generalize the well-known histogram methods~\cite{silverman1986density}. Simple probability functions called \emph{kernels} are placed at sample data points to approximate the underlying density function. 
        In the KDE framework, the choice of the kernel width influences the degree of smoothing of the estimated density function. Selecting the
    kernel width is known as the \emph{bandwidth selection problem}. The latter can be formulated simply as a bias-variance trade-off. Bandwidth selection is an extensively studied problem  (see~\cite{jones1996brief} for a brief review) with notable usages in signal processing, e.g.\ 
    mutual information estimation~\cite{moon1995estimation}. The 
    convergence of kernel density estimators for mixing dynamical systems was recently shown in~\cite{hang2018kernel}.
    
    The relation between kernel density estimation and the correlation sum is noted in~\cite{yu2000efficient} to estimate dynamical invariants in noisy situations.  More recently, Gaussian kernels estimators of the correlation integral are applied to estimate R\'enyi's entropies \cite{principe2010information, singh2011information, erdogmus2006linear}. 
    Here, KDE is used not to derive new estimators of nonlinear measures but rather as a framework providing a systematic rule to select the radius in computing nonlinear measures. 
    We show that the radius minimizing the relative error of the correlation sum estimator is equivalent to the bandwidth minimizing the Mean Integrated Squared Error (MISE) of a density estimator (Section~3.1). Therefore, we use a bandwidth selection method from KDE to derive a closed-form expression 
    for the optimal radius (Section~3.2) and define a ``meaningful range'' for the radius variable relatively to our optimum (Section~3.3). We conduct numerical experiments
    on well-known dynamical systems. First, we study the behavior of the correlation sum estimator in the ``meaningful range'' for signals of different 
    lengths and noise levels (Section~4). Then, we estimate the Kolmogorov-Sinai entropy of both simulated and real signals, using recurrence plots computed with an optimal radius (Section~5).

\section{Correlation sum and correlation dimension}
% \subsection{Correlation integral, correlation sum, correlation dimension}
    Let $(\mathcal{X}, \mathcal{A}, \mu, T)$ be a measure-preserving dynamical system with 
    $\mathcal{X} \subset \mathbb{R}^d$ and $\mu$ the invariant measure 
    (probability distribution in the phase space invariant upon the dynamics).
    The \emph{correlation integral} $c(r)$ is the mean probability to find a pair of points at two different time 
    $x,y \in \mathcal{X}$ arbitrarily close, such that the distance between $x$ and $y$ is less than a small radius parameter $r$~\cite{ott2002chaos, singh2011information}: 
    \begin{equation}
        \label{eq:corr_int}
        c(r) = {\mathbb P}\left((x, y) : \|x-y\|_p < r\right) = \int\limits_{x \in \mathcal{X}} \mu(B_r(x)) d\mu(x)
    \end{equation} 
    where $B_r(x) = \{y \in \mathcal{X} : \|x - y\|_p < r\}$ is the generalized $d$-dimensional ball in $L_p$ space,  with radius $r$ and center $x$.
    In practice, an estimator of the correlation integral  can be computed from a sample trajectory $x_i \in \mathbb{R}^d, 1\leq i \leq n$ \cite{grassberger1983characterization}: 
    \begin{align}
        \label{eq:corr_sum_asymp}
        C(r,n) = 
           \frac{1}{n^2} \sum\limits_{i,j=1}^n \Theta(r - \|x_i - x_j\|_p) \xrightarrow[n\to\infty]{} c(r)
    \end{align}
    where $\Theta$ is Heaviside step function and $C(r,n)$ is called the \emph{correlation sum} \cite{pesin2008dimension, grassberger1983characterization}. 
    For small values of $r$, the correlation integral grows as a power law: 
        \begin{align}
            \label{eq:powerlaw}
            c(r) \approx \text{const} \times r^{D_2}
        \end{align}
    The quantity $D_2$ is called the \emph{correlation dimension}.
  
\section{Kernel density estimation}
    \label{section:kde}
     A probability density function $f$ may be estimated 
    by placing \emph{smoothing kernels} at each sample point. 
    A smoothing kernel $K$ is defined as a valid probability density function, which satisfies \cite{silverman1986density}: 
    \begin{align}
        \int K(u) du = 1 && \forall u \in \mathbb{R}, K(u) \geq 0
    \end{align}
    Without loss of generality, we introduce a scaled version of the kernel with a $L_p$ norm and a scaling factor $h > 0$, 
    $K_h(u) = h^{-d}  K(u / h)$, which is a valid kernel when $K$ is a valid kernel. A simple kernel is the \emph{uniform} or \emph{boxcar} kernel, which
    remains constant over a domain:
    \begin{align}
        \label{eq:uniform_kernel}
        K_h(u) = \frac{1}{\tau_{p, d} h^d} \Theta(h - \|u\|_p)
    \end{align}
     where $\Theta$ is Heaviside step function and $\tau_{p, d}$ is the volume of the unit ball defined by the norm $p$ in a $d$-dimensional space (see Appendix~\ref{section:app_ball}).
    Given samples $x_i\in \mathbb{R}^d, 1 \leq i \leq n$, distributed  according to a density $f$, a kernel density estimator of $f$ is: 
    \begin{align}
        \label{eq:kdedef}
        \hat{f}_h(x) = n^{-1} \sum\limits_{i=1}^{n} K_h(x - x_i)
    \end{align}
    
     While kernel density estimators are consistent for i.i.d.\ samples, independence between consecutive samples cannot generally be assumed for dynamical systems. Hang ~\emph{et al.}~\cite{hang2018kernel} showed that kernel density estimators are also consistent for dynamical systems with mixing properties and weakly-continuous density function (more specifically $\mathcal{C}$-mixing systems with pointwise $\alpha$-H\"older controllable density, see~\cite{hang2018kernel} defs.\ 1 and 2).
     
     The \emph{bandwidth} parameter $h$  determines the ``width'' of the kernels and consequently 
      the degree of smoothing of the estimator. A plethora of methods exist to select the bandwidth parameter, 
      see \cite{jones1996brief}. Among existing bandwidth selection methods, minimizing the Asymptotic Mean Integrated Squared Error (AMISE), a Taylor expansion of the MISE of the estimator $\mathbb{E}[\int_{\mathbb{R}^d}(f(x) - \hat{f}(x))^2 dx]$, is appealing for 
      practical applications as it allows to derive a closed-form expression of 
      an approximately optimal bandwidth \cite{silverman1986density}:  
            \begin{equation}
                \label{eq:h_AMISE}
                h_{\text{AMISE}} = \left[\dfrac{W_1(K) \times d}{n  \times \left[W_2(K)\right]^{2} \times W_1(\nabla^2f)}\right]^{1/(d+4)} 
            \end{equation} 
            where the functionals $W_i$ are defined as $W_1(g) = \int_{\mathbb{R}^d} g^2(x) dx$ and
            $W_2(g) = \int_{\mathbb{R}^d} x_1^2 g(x) dx$, where $x_1$ is a scalar component of $x$. Reference rules \cite{silverman1986density, scott1979optimal} 
            can be easily obtained by replacing the unknown quantity $W_1(\nabla^2f)$ with the quantity computed using a reference distribution, generally a Gaussian distribution. 
    
\section{A reference rule for the optimal radius}
    To derive the expression of an optimal radius, we proceed as follows. First, we show that the radius minimizing the relative error of the correlation sum estimator is equivalent to the bandwidth minimizing the MISE of a particular density estimator.
    Second, we derive the closed-form expression of the radius minimizing the AMISE of the estimator. 
    Finally, we identify a meaningful range to select a variable radius. 
    
    \subsection{Criterion to select the radius}
           The correlation integral, $c(r) = \mathbb{E}_\mu\left[ \mu(B_r(x)) \right]$, is generally estimated by
           the correlation sum, $\hat{C}(r, n) =n^{-1} \sum_{i=1}^n
           \hat{\mu}(B_r(x_i))$ (\eqref{eq:corr_int} and \eqref{eq:corr_sum_asymp}). To obtain a good estimation of the correlation sum, we shall minimize the error between an estimator of the invariant measure of a ball , $\hat{\mu}(B_r(x))$, and the true quantity $\mu(B_r(x))$. However, minimizing such error is not sufficient to provide a good estimation for small $r$: the scale of the error decreases with $r$ and systematically leads to the trivial solution $r = 0$. Indeed, when $r$ decreases, the absolute error decreases while the relative error 
           is multiplied by a factor proportional to $1/r$ (\eqref{eq:powerlaw}) and consequently blows up. 
           Therefore, we want to find the radius  minimizing a relative error criterion on $\hat{\mu}(B_r(x))$. Let $\lambda$ be the Lebesgue measure, such that $\lambda(B_r)$ is the volume of a ball with radius $r$.
           We use the fact that $\mu(B_r(\cdot))$ is proportional to $r$ (see \cite{eckmann1985ergodic}, Section \romnum{5}.A.) and consequently that $\mu(B_r(\cdot)) \propto 
           \lambda(B_r)$  to simplify 
           the expression of the relative error and express the following local relative error: 
           \begin{align}
            \label{eq:relativeerror}
            L(r, x) = \mathbb{E} \left[ \left( \frac{\mu(B_r(x)) -\hat{\mu}(B_r(x))}{\lambda(B_r)}\right)^2 \right]
           \end{align}
           where the expectation is taken over samples used to construct the
           estimator. 
           Given a fixed $r$, $\mu(B_r(\cdot))$ is a bounded function on $\mathcal{X}$. We denote $\rho$ the normalized density of $\mu(B_r(\cdot))$, such that
           \begin{align}
             \rho(x) = \frac{\mu\left(B_r(x)\right)}{\int_\mathcal{X}\mu(B_r(x)) dx} = \frac{\mu\left(B_r(x)\right)}{\lambda(B_r)}
           \end{align}
           and, similarly, $\hat{\rho}_r(x) = \hat{\mu}(B_r(x)) / \lambda(B_r)$ the estimator of the normalized density 
           $\rho$. After replacing in \eqref{eq:relativeerror}, we obtain: 
           \begin{align}
            \label{eq:localdstyerr}
            L(r, x) =\mathbb{E} \left[ \left(\rho(x) - \hat{\rho}_r(x)  \right)^2 \right]
           \end{align}
           Then, integrating \eqref{eq:localdstyerr} over possible values of $x$ gives a global criterion to select the optimal radius $r_\text{opt}$:
           \begin{align}
            \label{eq:roptcrit}
            r_\text{opt} &= \argmin_r \mathcal{L}(r)
           \end{align} 
           where
            \begin{align}\mathcal{L}(r) &= 
                \label{eq:miseropt}
                  \int_\mathcal{X} \mathbb{E}\left[\left(\rho(x) - \hat{\rho}_r(x)  \right)^2 \right] dx
           \end{align} 
           With simple manipulations, we see that \eqref{eq:miseropt} is indeed the MISE between the estimator of the normalized density $\hat{\rho}_r$ and the true 
           normalized density $\rho$ \cite{silverman1986density}. Finally, $\hat{\rho}_r$ can be identified 
           by replacing $\hat{\mu}(B_r(x))$ with $\hat{\rho}_r$ in the expression of the correlation sum estimator (\eqref{eq:corr_sum_asymp}): 
           \begin{align}
               C(r, n) &= \frac{\lambda(B_r)}{n} \sum_{i = 1}^n \hat{\rho}_r(x_i) 
                     = \frac{1}{n^2} \sum_{i,j  = 1}^n \Theta(r - \|x_i - x_j\|) \nonumber 
            \end{align}
                yielding 
            \begin{align}
            \hat{\rho}_r(x) &= \frac{1}{n \lambda(B_r)} \sum_{i=1}^n   \Theta(r - \|x - x_i\|)
           \end{align}
           As $\lambda(B^{p,d}_r) = \lambda(B^{p,d}_1) r^d$ (\ref{section:app_ball}), 
           we observe that $\hat{\rho}_r$ is a kernel density estimator with a uniform kernel 
           and a bandwidth parameter $r$ (\eqref{eq:uniform_kernel}).  Hence, it follows from \eqref{eq:roptcrit} that 
           the bandwidth minimizing the MISE of the estimator $\hat{\rho}_r$ can provide a good approximation of the optimal radius $r_\text{opt}$ minimizing the relative error
           on the correlation sum estimator. 
           Moreover, the AMISE method (\eqref{eq:h_AMISE}) can be used to approximate $r_\text{opt}$ with a simple, closed-form expression that resembles to the empirical rules currently used. 
           In the next section, we use the AMISE minimization 
           method to derive a reference rule for the optimal radius.  
           
    \subsection{Derivation of a reference rule for the optimal radius}
            As presented in \autoref{section:kde}, a Taylor expansion of the MISE can be used to derive a closed-form expression  of the optimal bandwidth for an estimator. A particular interest for this method is motivated by the possibility of deriving a closed-form expression of the optimal bandwidth. We use a reference Gaussian distribution in \eqref{eq:h_AMISE} and derive the expressions for $W_1(K)$ and $W_2(K)$ for the uniform kernel (see Appendix~\ref{section:app_refrule}). Then, substituting these expressions into~\eqref{eq:h_AMISE} gives the main result of the paper: a \emph{reference rule radius} $r_{\text{opt}}$ defined as
            \begin{align}
                \label{eq:refrule}
                r_\text{opt} &= \alpha_{p, d} \times \hat{s} \times  n^{-1/(d+4)}
                % h_{\text{opt}} &= \alpha_{p, d} \sigma n^{-1/(d+4)}
            \end{align}
            where $\alpha_{p, d}$, depending on the norm and dimension, rescales $r_\text{opt}$; $\hat{s}$ is an estimate of the spread of data; and $n$ is the length of the 
            trajectory in phase space. 
            \begin{remark}
            \label{remark:rec_series_length}
            In practice, the phase space is reconstructed using a time delay embedding procedure (according to Takens theorem \cite{takens1981detecting}) ; hence, if 
            $N$ denotes the length of the univariate time series, $d$ the embedding dimension and $\tau$ the delay, the length of the trajectory in reconstructed phase
            space is $n = N - (d - 1) \tau$.
            \end{remark}
            
            \subsubsection{Estimation of the spread $\hat{s}$}
            A first choice for the spread $\hat{s}$ is the average marginal sample standard 
            deviation, defined by $\hat{\sigma} = \sqrt{d^{-1} \sum_i S_{i,i}}$, 
            $S\in \mathbb{R}^{d\times d}$ is the sample covariance matrix. When the $d$-dimensional sample is constructed from an univariate time series using delay embedding,  components on the diagonal of the sample covariance matrix are equal: 
            $\hat{\sigma}$ is then the sample standard deviation of the time series.  Alternatively,
            as the interquartile range $\text{IQR}$ is a good alternative to standard deviation for non-Gaussian data (see \cite{silverman1986density} for discussion), a common choice for $\hat{s}$ is: 
            \begin{align}
                \hat{s} = \min\left(\hat{\sigma}, \dfrac{\mathrm{IQR}}{1.34}\right).
            \end{align}
            
            \subsubsection{Derivation of the reference factor $\alpha_{p, d}$}
            The expression for the 1-dimensional reference factor is relatively straightforward: $\alpha_{p, 1} = (12\sqrt{\pi})^{1/5} \approx 1.843$.
            The general closed-form expression for $\alpha_{p, d}$ is more complex (see Appendix~\ref{section:app_refrule}, \eqref{eq:app_genexpr}); 
            however, the expression can be simplified for common norms (Appendix~\ref{section:app_splfy}): 
            \begin{align}
                    \alpha_{1, d} &=  \left[\left(d+2\right)!\, (d+1)  (\sqrt{\pi})^d  \right]^{1/(d+4)} 
                \label{eq:refl1} \\
                \alpha_{2, d} &= 2 \times \left[ \frac{\Gamma\left(\frac{d}{2} + 2\right)}{2} \right]^{1/(d+4)} 
                \label{eq:refl2} \\
                     \alpha_{\infty, d} &= \left[ \frac{36 (\sqrt{\pi})^d }{d+2} \right]^{1/(d+4)}
                \label{eq:reflinf}
            \end{align}
            Moreover, $\alpha_{p, d}$ is to be computed only once for common 
            dimensions and norms. Hence, we report in Table \ref{table:alpha_values} some values of $\alpha_{p, d}$ that can be used in~\eqref{eq:refrule}.
            \begin{table}[h]
                % % \vspace{-0.5em}
                \begin{center} 
                \begin{tabular}{cl|ccc}
                    \multicolumn{2}{l}{\multirow{2}{*}{$\alpha_{p, d}$}} & \multicolumn{3}{|c}{$p$} \\
                    \multicolumn{2}{l|}{}                  & 1 & 2 & $\infty$ \\
                    \hline
                    \multirow{5}{*}{$d$}       & 1        & 1.843 & 1.843 & 1.843   \\ 
                                               & 2        & 2.468 & 2.000 & 1.745   \\ 
                                               & 3        & 3.087 & 2.150 & 1.694   \\   
                                               & 4        & 3.705 & 2.294 & 1.666   \\
                                               & 5        & 4.325 & 2.432 & 1.649  
                \end{tabular}
                % % \vspace{-1.5em}
                \end{center}
                        \caption{Rounded values of the coefficient $\alpha_{p, d}$ for common norms and dimensions. 
                        The values can directly be used in reference rule radius, $r_\text{opt} = \alpha_{p, d} \times \min\left(\hat{\sigma}, \mathrm{IQR}/1.34\right) \times  n^{-1/(d+4)}$.}
                        \label{table:alpha_values}
                % \vspace{-1.5em}
            \end{table}
            
    \subsection{Identification of a meaningful range for a variable radius}
             \label{section:range}
             As discussed above, some nonlinear indices 
        require selecting the range of radius values in which the quantity is estimated.
        For instance, this applies to nonlinear indices quantifying a scaling exponent of the form $\lim\limits_{r\to 0} \frac{\log \nu(B_r)}{\log r}$ 
        (with $\nu$ a (with nu a probability distribution in the phase space),
        as often encountered in the chaotic systems literature (see e.g. \cite{pesin2008dimension,ott2002chaos}). 
        In practice, the limit ${r\to0}$ is generally intractable, and 
        estimations of $\nu$ for small $r$ are highly variable due to poor statistics \cite{eckmann1985ergodic}. 
        On the other hand, at a certain point, large $r$ will not capture the 
        desired scaling effect. Hence, there is a range of values which must be selected to support a good estimation of the nonlinear measure. 
        Here, we introduce our arguments to guide the selection of a meaningful range for a variable radius. 
        
        The AMISE can be expanded in an integrated squared bias and integrated variance of the density estimator, 
        giving the following expressions of bias and variance as functions of 
        $r$ \cite{silverman1986density}: 
        \begin{align}
                \label{eq:bias}
                \mathrm{bias}(r) &\simeq \frac{r^2}{2}  W_2(K) \nabla^2 \rho(x) \\
                \label{eq:var}
                \mathrm{var}(r) &\simeq \frac{W_1(K)}{n r^{d}} \rho(x) 
        \end{align}
        The behavior of the relative error with $r$ can be understood from \eqref{eq:bias} and \eqref{eq:var}:
        the bias is proportional to $r$ whereas the variance is inversely proportional to $r$. As $r_\text{opt}$ minimizes 
        the AMISE, the bias contribution increases with $r$ while the variance decreases with $r$.  
        However, the bias term only depends on $r$ whereas the variance term decreases when the number of points increases.  
        This observation --considering that usually the bandwidth minimizing the AMISE is too large, suggests
        selecting the optimal radius $r_\text{opt}$ as the upper bound for the meaningful range. We introduce a 
        range parameter $0 < \beta < 1$ to select the lower bound as a fraction of $r_\text{opt}$, such that the radius values lie within the range: 
        \begin{align}
            \mathcal{R} = \left[\beta r_\text{opt}, r_\text{opt}\right]
        \end{align}
        Due to the relations \eqref{eq:bias} and \eqref{eq:var}, we argue that the value of $\beta$ shall be decreased when increasing the number of points.

\section{Estimation of the correlation dimension}
    \label{section:corrdim}
    In the following, we investigate the behavior of the Grassberger and Proccacia algorithm for the estimation of 
    the correlation dimension. We compare the spread and bias of estimations in the full range of available scales with estimations
     in the meaningful range derived in \autoref{section:range}. 
    \subsection{The Grassberger and Proccacia algorithm}
        
    The correlation dimension $D_2$ can be expressed as: 
        \begin{align}
            D_2 = \lim \limits_{r \to 0} \dfrac{\log c(r)}{\log r} =\lim \limits_{r \to 0} \lim \limits_{n \to \infty}
            \dfrac{\log C(r,n)}{\log r}
        \end{align}
        The Grassberger and Proccacia algorithm \cite{grassberger1983characterization} for the empirical estimation of the correlation dimension consists in computing the correlation sum for different values of $r$ and plotting $\log C(r, n)$ versus $\log r$. The slope of the linear region in this logarithmic plot provide the desired estimation of the correlation dimension $D_2$ \cite{ott2002chaos}. Similarly to the original paper \cite{grassberger1983characterization}, we use linear regression to estimate the slope. 

    \subsection{Procedure for generating reconstructed trajectories}
    \label{section:num_exp_procedure}
           
    \begin{figure}[h!]
    % \vspace{-0.5em}
    \centering
    
    \begin{subfigure}{0.9\textwidth}
        \centering
        \includegraphics[width=0.9\linewidth, trim={0 0 0 2em}, clip]{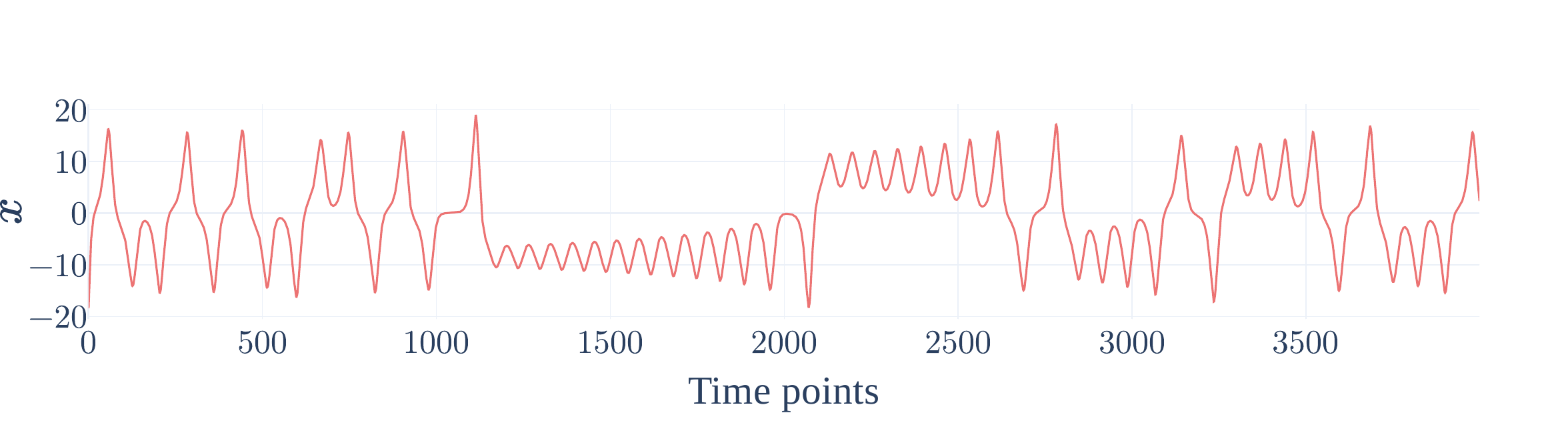}
    \end{subfigure}
    \hfill
    \begin{subfigure}[t]{0.03\textwidth}
        \caption{\label{fig:sample_series_lorenz}}
    \end{subfigure}
    
    \begin{subfigure}{0.9\textwidth}
        \centering
        \includegraphics[width=0.9\linewidth, trim={0 0 0 4em}, clip]{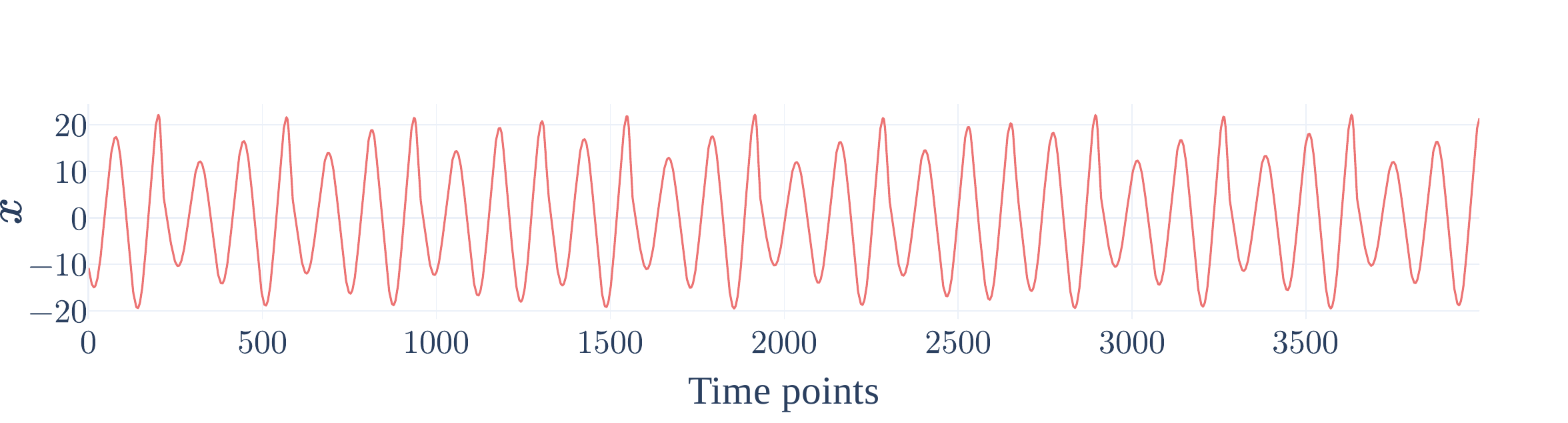}
    \end{subfigure}
    \hfill
    \begin{subfigure}[t]{0.03\textwidth}
        \caption{\label{fig:sample_series_rossler}}
    \end{subfigure}
    
    \begin{subfigure}{0.9\textwidth}
        \centering
        \includegraphics[width=0.9\linewidth, trim={0 0 0 4em}, clip]{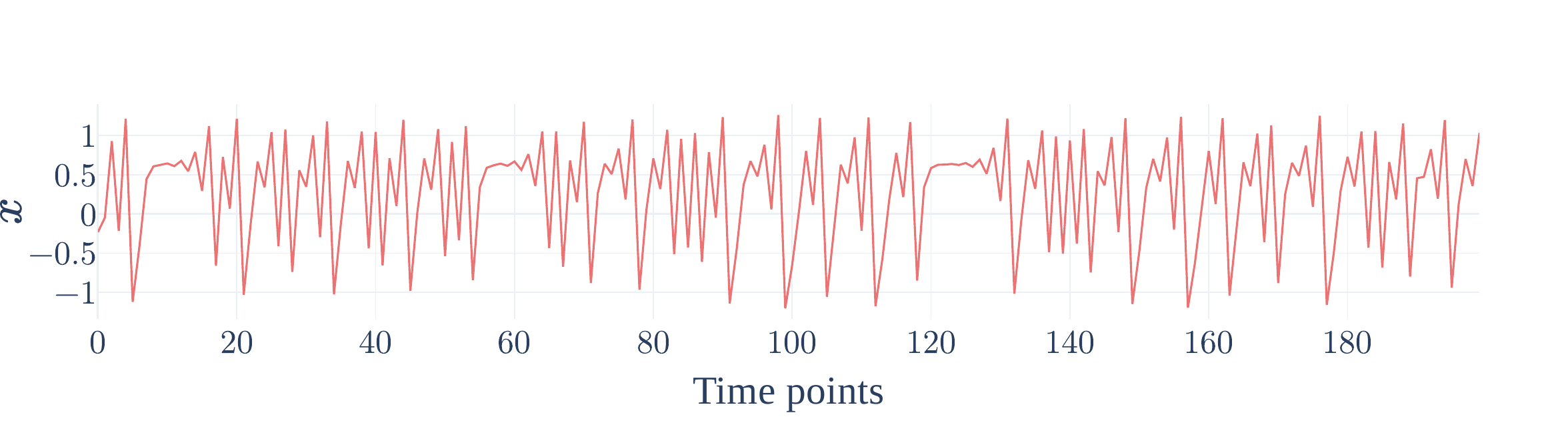}
    \end{subfigure}
    \hfill
    \begin{subfigure}[t]{0.03\textwidth}
        \caption{\label{fig:sample_series_henonmap}}
    \end{subfigure}
    \caption{Sample series for the Lorenz attractor (\subref{fig:sample_series_lorenz}), the R\"ossler attractor (\subref{fig:sample_series_rossler}) and the H\'enon map (\subref{fig:sample_series_henonmap}). Series from the $x$ coordinates were systematically used to reconstruct the trajectory using Takens 
    delay embedding.}
    \label{fig:sample_series}
\end{figure}
    We conducted numerical experiments on the Lorenz system \cite{lorenz1963deterministic} 
    ($\sigma=10$, $\beta=\frac{8}{3}$, $\rho=28$, $dt = 0.01$), 
    the R\"ossler system \cite{rossler1976equation} ($a=0.1$, $b=0.1$, $c=14$, $dt=0.05$) and the H\'enon map \cite{henon1976two} ($a=1.4$, 
    $b=0.3$). 
    We apply the following procedure to generate random time series of different length. 
    After drawing a random initial state, we generate time series for all systems -- 
    using a Runge-Kutta 4/5 method for Lorenz and R\"ossler -- such that the length of the time series is $N$ after removing transients (sample series are presented in \autoref{fig:sample_series}).  
    Then, the trajectory is reconstructed using Takens delay embedding (series from the $x$ coordinates were systematically used). 
    The original system dimension is used as embedding dimension $d$. 
    The time delay parameter $\tau$ is set to $1$ for the H\'enon map and selected
     as the first minimum of the time-delayed mutual information function~\cite{fraser1986independent} for the Lorenz and R\"ossler systems. 
     Please note that the length of the reconstructed 
     trajectories, $n = N - (d -1) \tau$, is used to compute the optimal radius using \eqref{eq:refrule} (see Remark~\ref{remark:rec_series_length}).
    \begin{remark}
    Here, we assume that the delay and embedding dimension are correctly selected as a bad phase space reconstruction 
    deteriorates the $\log C(r, n)$ versus $\log r$ plot \cite{kantz2004nonlinear}. In practice, this embedding problem can be efficiently addressed 
    as a plethora of methods exist to select the delay and the embedding dimension (see for instance \cite{kantz2004nonlinear}, Ch.~3.3 and Ch.~9.2). 
    \end{remark}
    
    \subsection{Numerical results for the radius range}
        \label{section:corrdim_results_range}
 
\begin{figure}[ht!]
    \centering
    \begin{subfigure}{0.9\textwidth}
        \centering
        \includegraphics[width=0.8\linewidth, trim={0 0 0 4em}, clip]{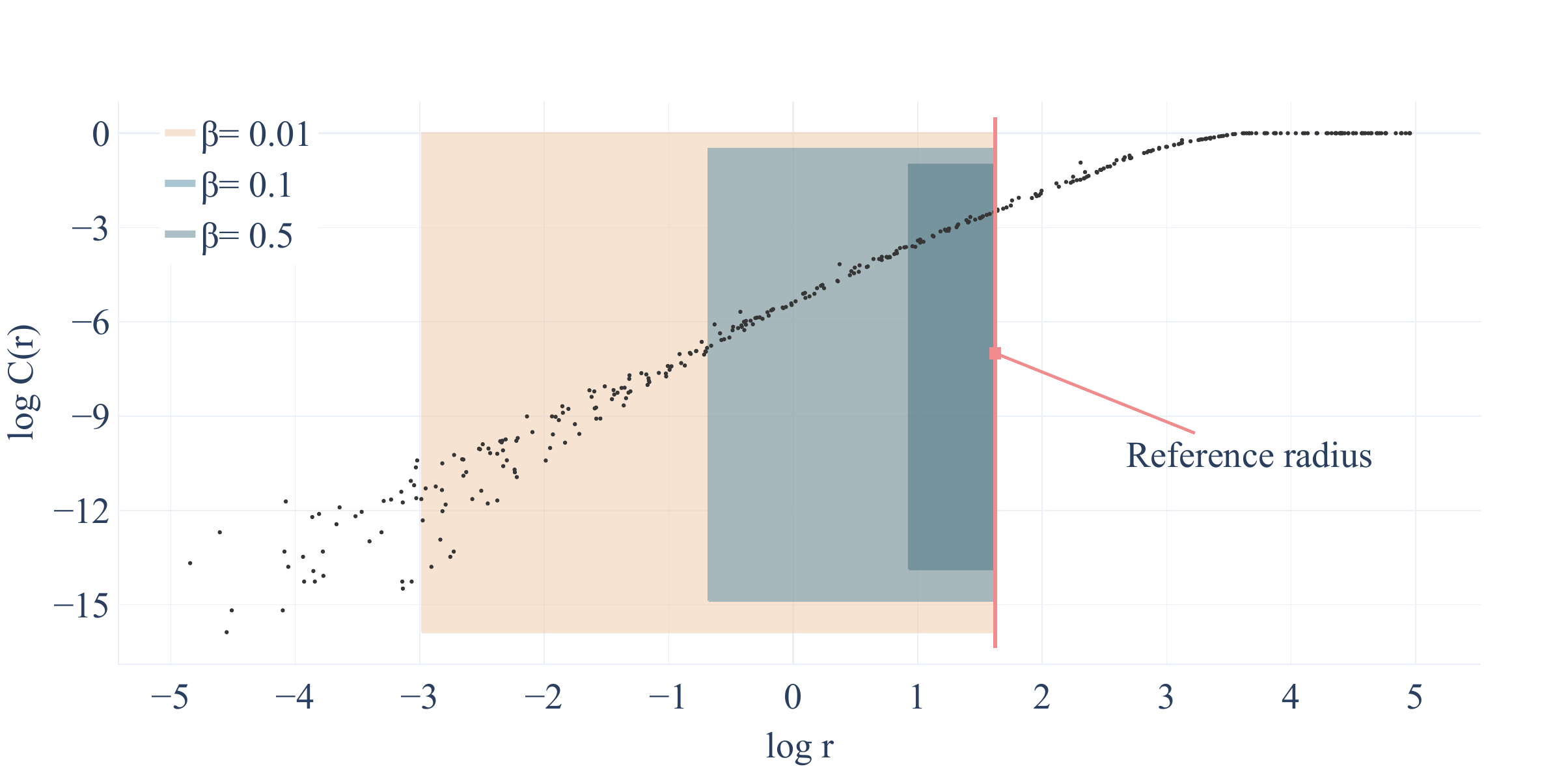}
    \end{subfigure}
    \hfill
    \begin{subfigure}[t]{0.03\textwidth}
        \caption{\label{fig:plot_corrdim_beta_Lorenz}}
    \end{subfigure}
    \begin{subfigure}{0.9\textwidth}
            \centering
        \includegraphics[width=0.8\linewidth, trim={0 0 0 4em}, clip]{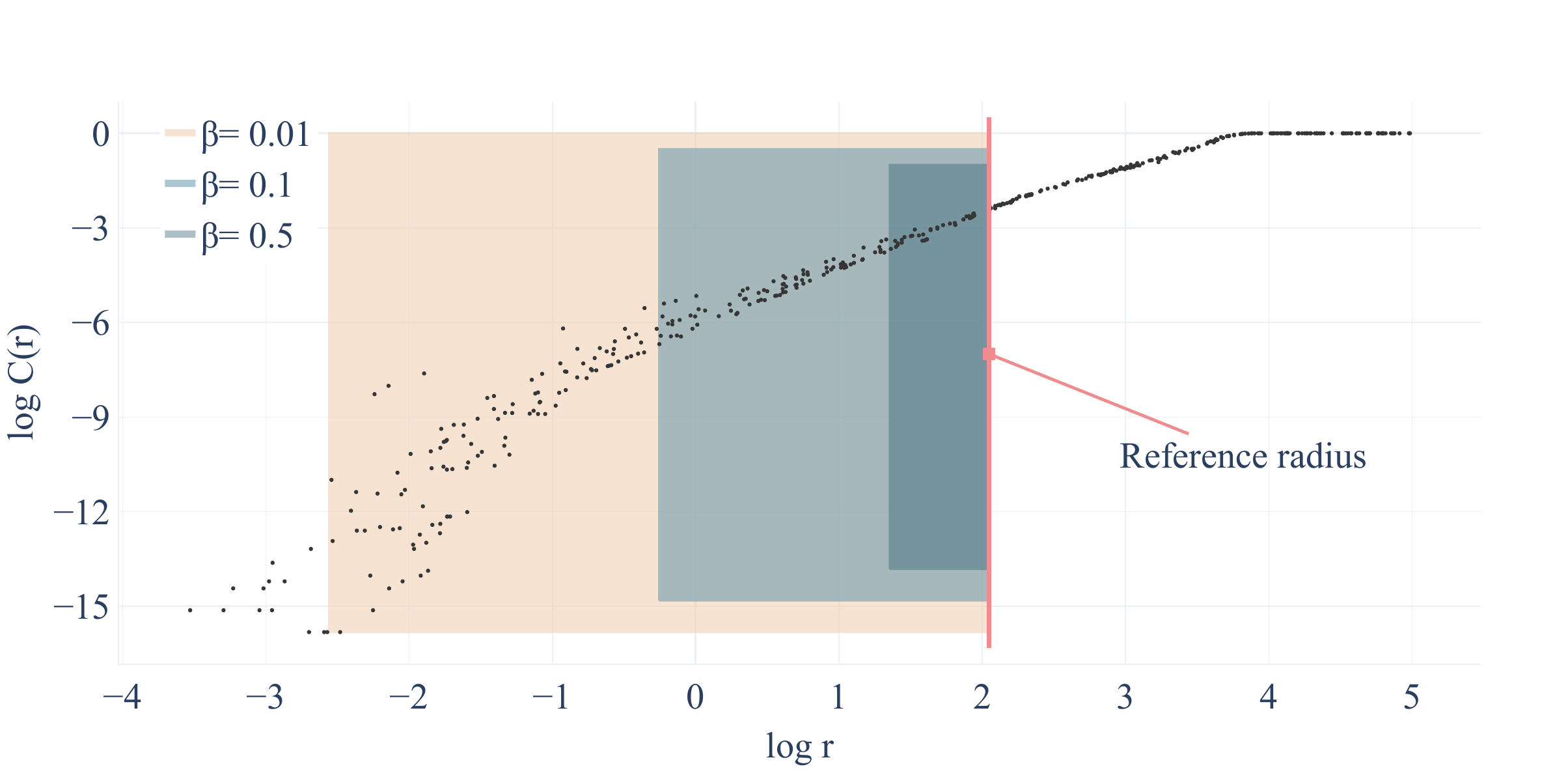}
    \end{subfigure}
    \hfill
    \begin{subfigure}[t]{0.03\textwidth}
        \caption{\label{fig:plot_corrdim_beta_Rossler}}
    \end{subfigure}
    \begin{subfigure}{0.9\textwidth}
        \centering
        \includegraphics[width=0.8\linewidth, trim={0 0 0 4em}, clip]{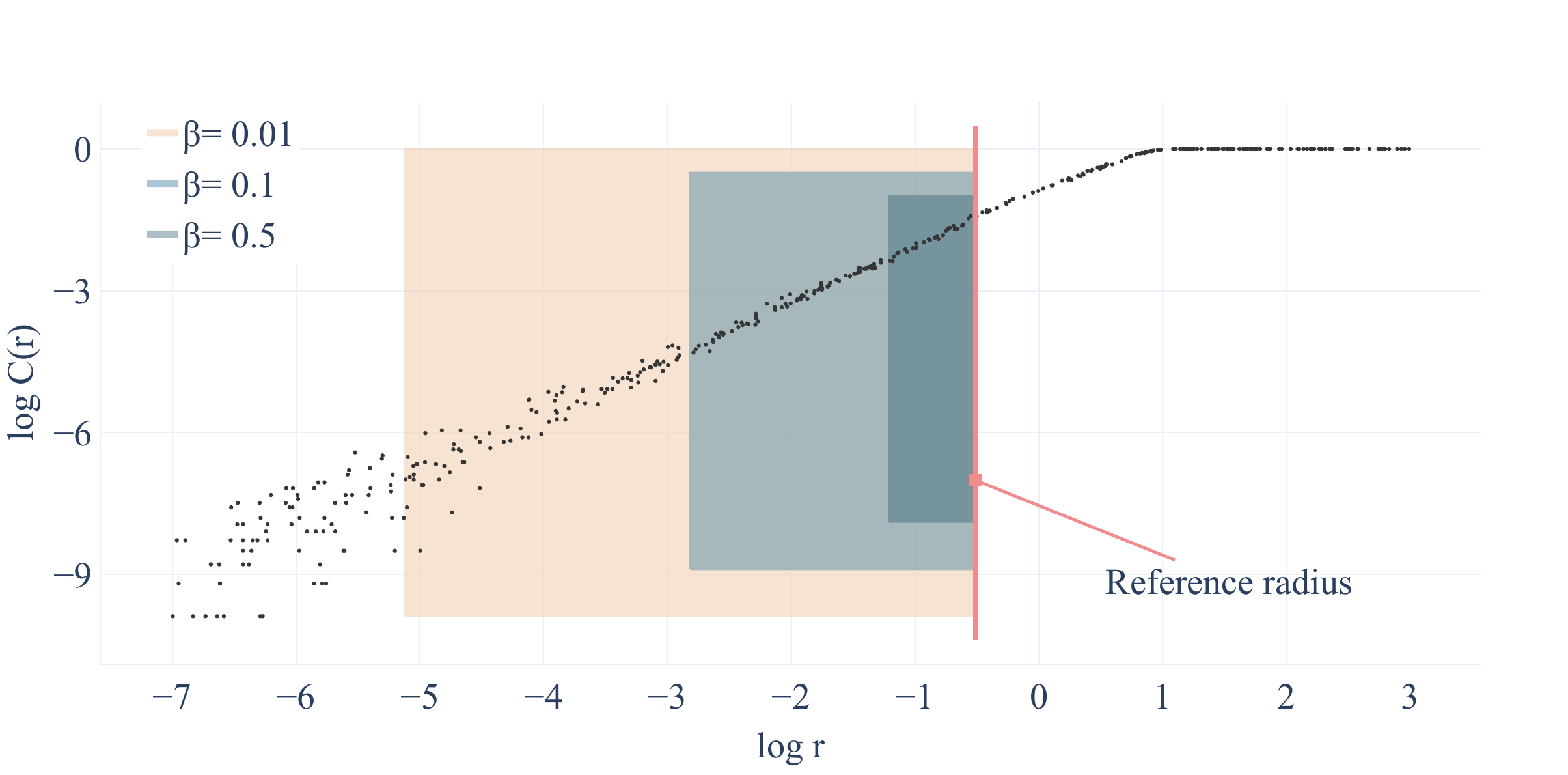}
    \end{subfigure}
    \hfill
    \begin{subfigure}[t]{0.03\textwidth}
        \caption{\label{fig:plot_corrdim_beta_HenonMap}}
    \end{subfigure}
        \caption{Log-log plot of correlation sums versus the radius used to estimate the correlation dimension for Lorenz (\subref{fig:plot_corrdim_beta_Lorenz}), 
        R\"ossler (\subref{fig:plot_corrdim_beta_Rossler}) and H\'enon (\subref{fig:plot_corrdim_beta_HenonMap}) systems. The dots correspond to estimates of the correlation sum
        at random values of radius for trajectories integrated from random initial states. 
        The colored regions show different ranges defined 
        by $\left[\beta r_\text{opt}, r_\text{opt}\right]$, with $\beta = 0.01$ (in beige), $0.1$ (in light blue) or $0.5$ (in dark blue).}
        \label{fig:corr_dim}
    \end{figure}

    Here, we visualize the meaningful range on the log-log plot of the correlation sum versus the radius. 
    We first generate 100 series for each system (4000 points 
    for  R\"ossler and Lorenz attractor, 200 points for H\'enon map, respectively), select 25 random values of radius and compute the corresponding correlation sums. 
    We overlay the average value of $r_\text{opt}$ and the ranges with arbitrary values $\beta \in \{0.01, 0.1, 0.5\}$ on the plot of $\log C(r,n)$ vs $\log r$. Results are presented in \autoref{fig:corr_dim}. We observe 
    that the spread of the correlation sum over the runs is low at the location of reference radius and increases when the radius is decreased. Hence, smaller values of $\beta$ 
    likely lead to higher variance estimations.
    
    On \autoref{fig:plot_corrdim_beta_Rossler}, a knee is present around a value $\log(r_\text{knee})\simeq -1$, such that the
    slope $a_\text{left}$ in a left range $[r_0, r_\text{knee}]$ is higher than the slope $a_\text{right}$ in right range $[r_\text{knee}, r_1]$. In practice,
    a knee may appear from the superposition of signals from non-interacting subsystems with different amplitude
    \cite{eckmann1985ergodic}. 
    In this situation, $a_\text{left}$ characterizes the two subsystems while $a_\text{right}$ corresponds only to the 
    system with the largest signal amplitude. Consequently, a careful analysis of the plot of $\log C(r,n)$ vs $\log r$ 
    might be necessary to select a range capturing the desired properties of systems under study.
    
        \subsection{Influence of the time series length}
\begin{figure}[tb!]
    % \vspace{-0.5em}
        \centering      
        \begin{subfigure}{0.9\textwidth}
            \centering
            \includegraphics[width=.9\linewidth, trim={0 0 0 2em}, clip]{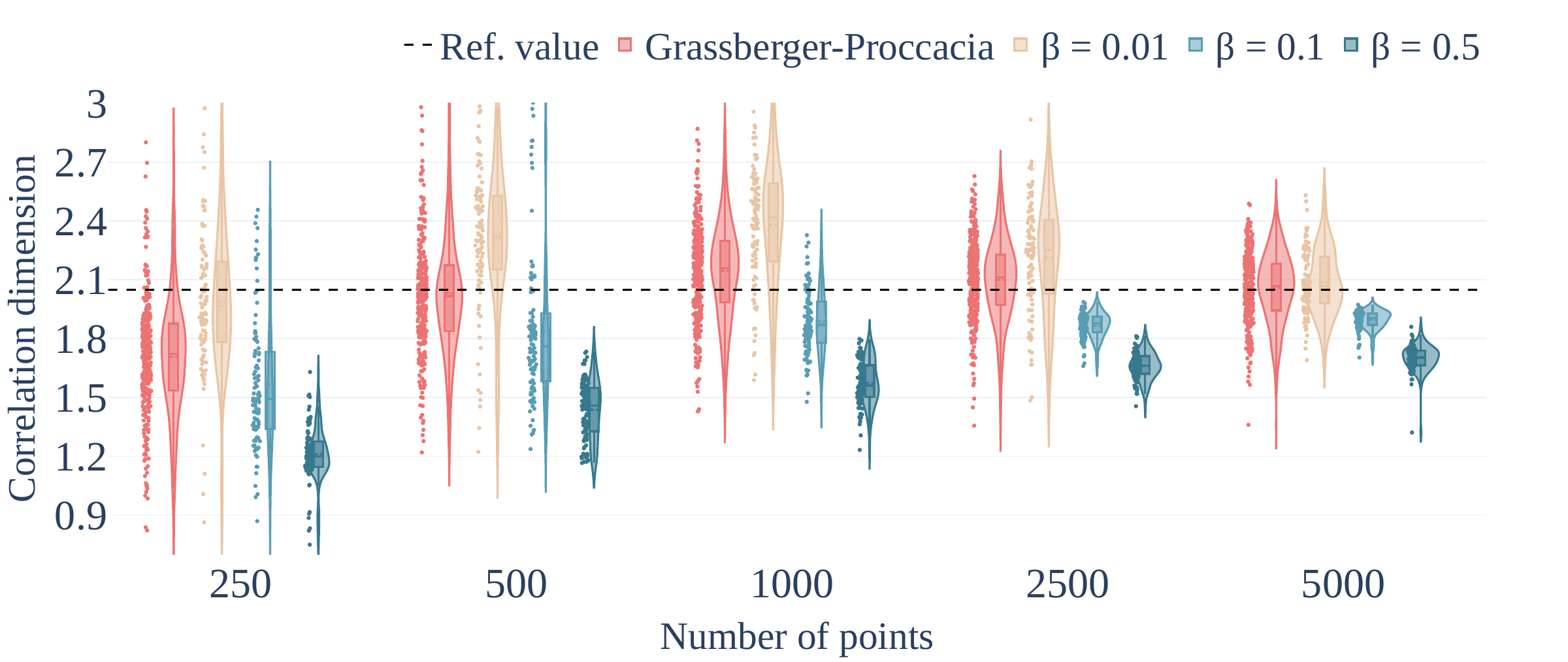}
        \end{subfigure}
        \hfill
        \begin{subfigure}[t]{0.03\textwidth}
            \caption{\label{fig:d2violins_beta_Lorenz}}
        \end{subfigure}
        
        \begin{subfigure}{0.9\textwidth}
        \centering
            \includegraphics[width=.9\linewidth, trim={0 0 0 4em}, clip]{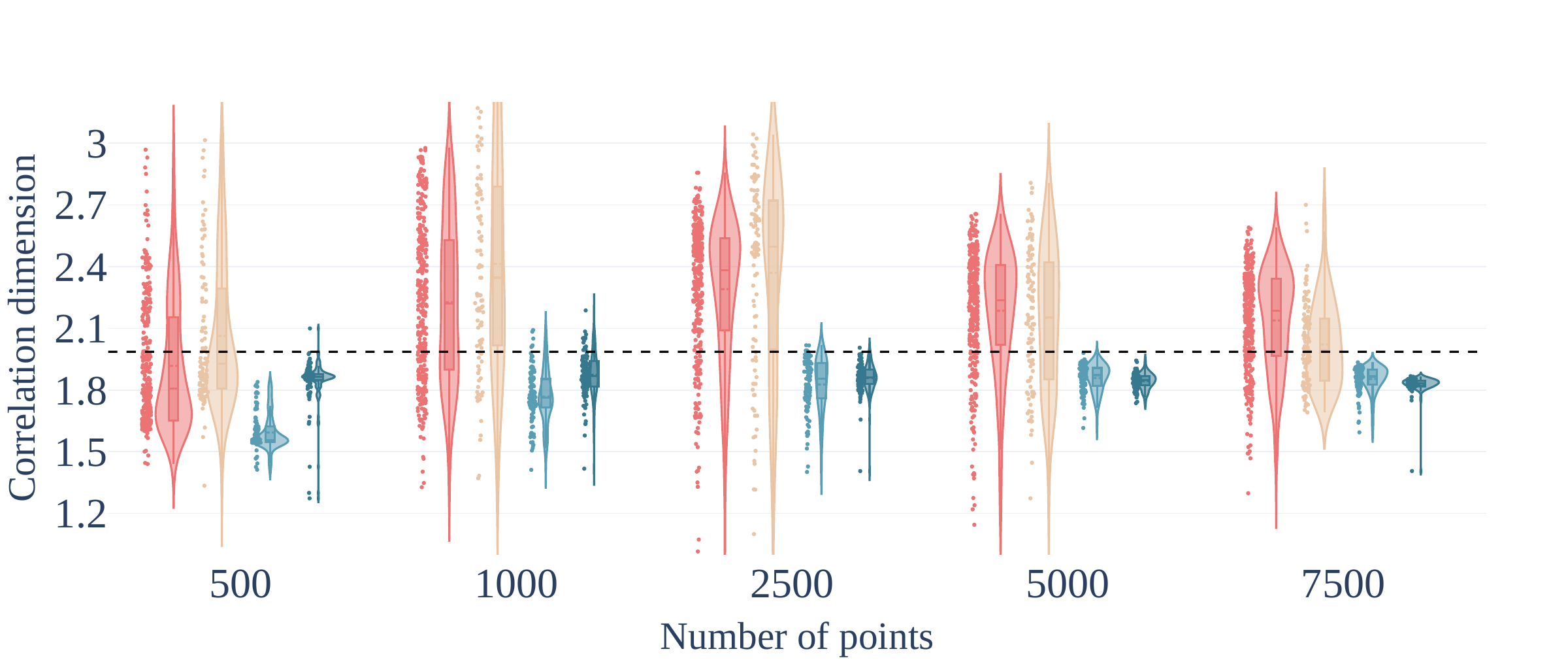}
        \end{subfigure}
        \hfill
        \begin{subfigure}{0.03\textwidth}
            \caption{\label{fig:d2violins_beta_Rossler}}
        \end{subfigure}
        
        \begin{subfigure}{0.9\textwidth}
        \centering
            \includegraphics[width=.9\linewidth, trim={0 0 0 4em}, clip]{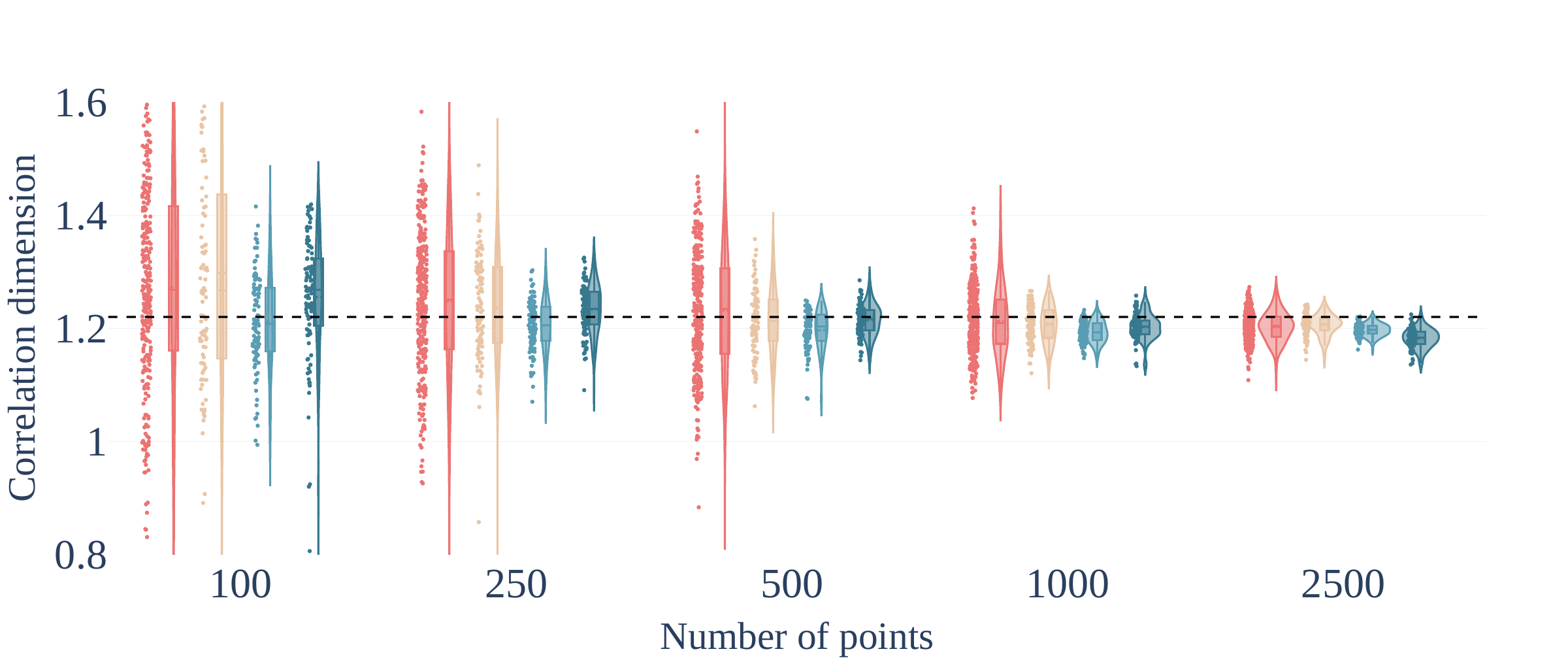}
        \end{subfigure}
        \hfill
        \begin{subfigure}{0.03\textwidth}
            \caption{\label{fig:d2violins_beta_HenonMap}}
        \end{subfigure}
        % \includegraphics[width=\textwidth]{./plot_d2_violins_beta.pdf}
    % \vspace{-0.5em}
        \caption{Influence of the number of points on the estimation of the correlation dimension from Lorenz (\subref{fig:d2violins_beta_Lorenz}), R\"ossler (\subref{fig:d2violins_beta_Rossler})
        and H\'enon (\subref{fig:d2violins_beta_HenonMap}) systems.
        We compare the original version of the Grassberger and Proccacia algorithm (red) with
        an estimation of the slope in the range $\left[\beta r_\text{opt}, r_\text{opt}\right]$, with $\beta = 0.01$ (beige), $\beta=0.1$ (light blue), $\beta = 0.5$ (dark blue).}
        \label{fig:corr_dim_violin}
    % \vspace{-0.5em}
    \end{figure}

    Using the procedure described in Section \ref{section:num_exp_procedure}, we generate $100$ trajectories for each length: 
    \begin{enumerate}[label=(\alph*)]
        \item $N=250, 500, 1000, 2500, 5000$ for the Lorenz system,
        \item $N=500, 1000, 2500, 5000, 7500$ for the R\"ossler system,
        \item $N=100,250,500,1000,2500$ for the H\'enon map. 
    \end{enumerate}
     \begin{remark}
        Notice that the discrepancies for the number of points used for the three systems can be justified
        by the resulting trajectories after time delay embedding. Indeed, when the number of points is too low, the reconstructed trajectories 
        cannot properly reflect the dynamics nor the correct dimension of the attractor. For instance, in our experiments this was the case 
        for time series of 100 points for the R\"ossler system.
     \end{remark}
     We computed correlations sums for $20$ values of $r$ ranging between $10^{-8}$ and $2 \sigma$, where $\sigma$ is the sample 
    standard deviation. 
    We compared the estimation using 
    the Grassberger and Proccacia algorithm on the entire curve (the plateau on the right was omitted) with the estimation 
    for $20$ values of $r$ in the meaningful range $(\beta r_\text{opt}, r_\text{opt})$, for values of $\beta \in \{0.01, 0.1, 0.5\}$. We show in \autoref{fig:corr_dim_violin} a violin plot of the estimated values depending on the duration of the
    time series and the range used to estimate the dimension. We compare our estimations with the values of correlation dimension reported by Sprott and Rowlands~\cite{sprott2001improved} for much longer series ($2.049 \pm 0.096$ for Lorenz attractor, $1.986 \pm 0.078$ for R\"ossler attractor, $1.220 \pm 0.036$ for H\'enon map). 
  
    Overall, we observe that the spread of the estimations decreases with increasing $\beta$. 
    With $\beta = 0.01$ (beige) the result is almost similar to the original version of the Grassberger and Proccacia algorithm (red). 
    In contrast, estimations for larger values of $\beta$ are more localized, but around values of dimension further apart from the reference 
    dimension. We observe that the number of points affects significantly the variance of the estimations for larger values of $\beta$. 
    However, for both R\"ossler and H\'enon attractors, the range parameter $\beta = 0.1$ (light blue) gives estimations with lower variance and bias compared to $\beta = 0.5$ (dark blue). This suggests that the range must be selected sufficiently large to provide a proper support for dimension estimation. 
    Moreover, although the bias of the Grassberger-Proccacia algorithm is low in this setup, a single dimension estimate can be far from the true 
    dimension. Therefore, one can favor a smaller range for $r$ to reliably estimate a quantity slightly lower than the true dimension.

    We found qualitatively similar results for Lorenz and R\"ossler 
    attractors when series of different length are obtained by downsampling an original series of fixed length (results not shown).  
    
    \subsection{Influence of observational white noise}

        Finally, we investigate the influence of observational noise on the estimation of the correlation dimension in 
        the different ranges. Observational noise is ubiquitous in practical applications and creates a knee on the plot of $\log C(r,n)$ versus $\log r$, with a dimension at the left of the knee equal to the embedding dimension 
        (see~\cite{eckmann1985ergodic,grassberger2004measuring}).
        Hence, the range  must be selected at the right of the knee to provide good estimations of the dimension.
        
\begin{figure}[h!]
    % \vspace{-0.5em}
        \begin{subfigure}{0.93\textwidth}
            \centering
            \includegraphics[width=.95\linewidth, trim={0 0 0 2em}, clip]{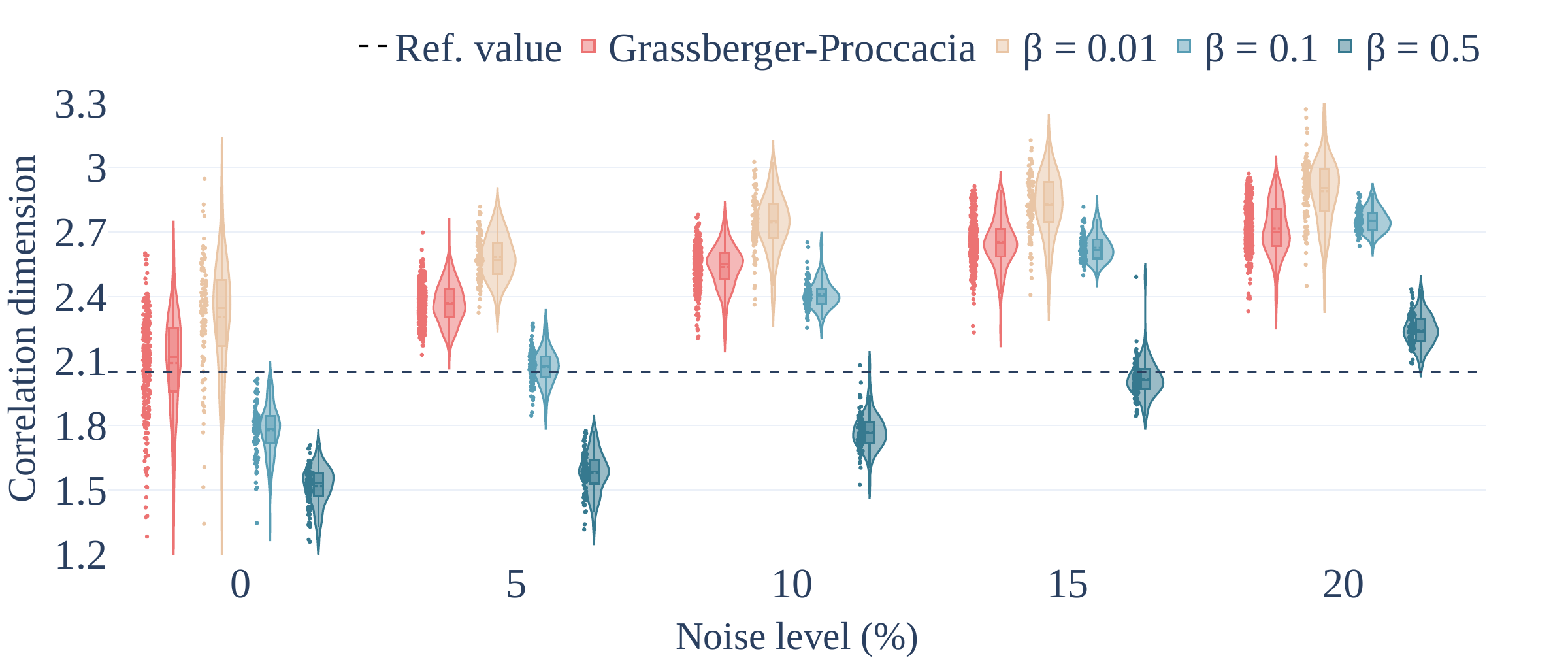}
        \end{subfigure}
        \hfill
        \begin{subfigure}{0.03\textwidth}
            \caption{\label{fig:d2violins_beta_noise_Lorenz}}
        \end{subfigure}
        \begin{subfigure}{0.93\textwidth}
        \centering
            \includegraphics[width=.95\linewidth, trim={0 0 0 4em}, clip]{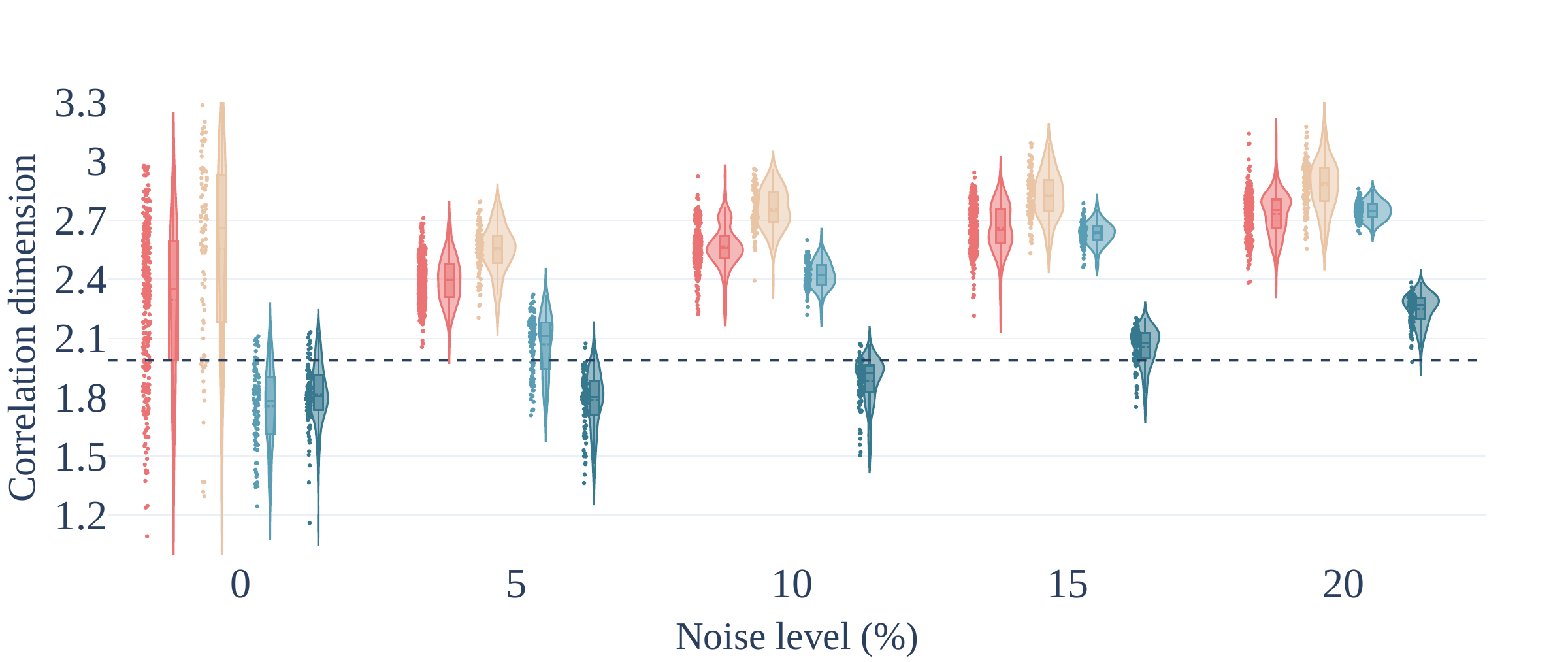}
        \end{subfigure}
        \hfill
        \begin{subfigure}{0.03\textwidth}
            \caption{\label{fig:d2violins_beta_noise_Rossler}}
        \end{subfigure}
        \begin{subfigure}{0.93\textwidth}
        \centering
            \includegraphics[width=.95\linewidth, trim={0 0 0 4em}, clip]{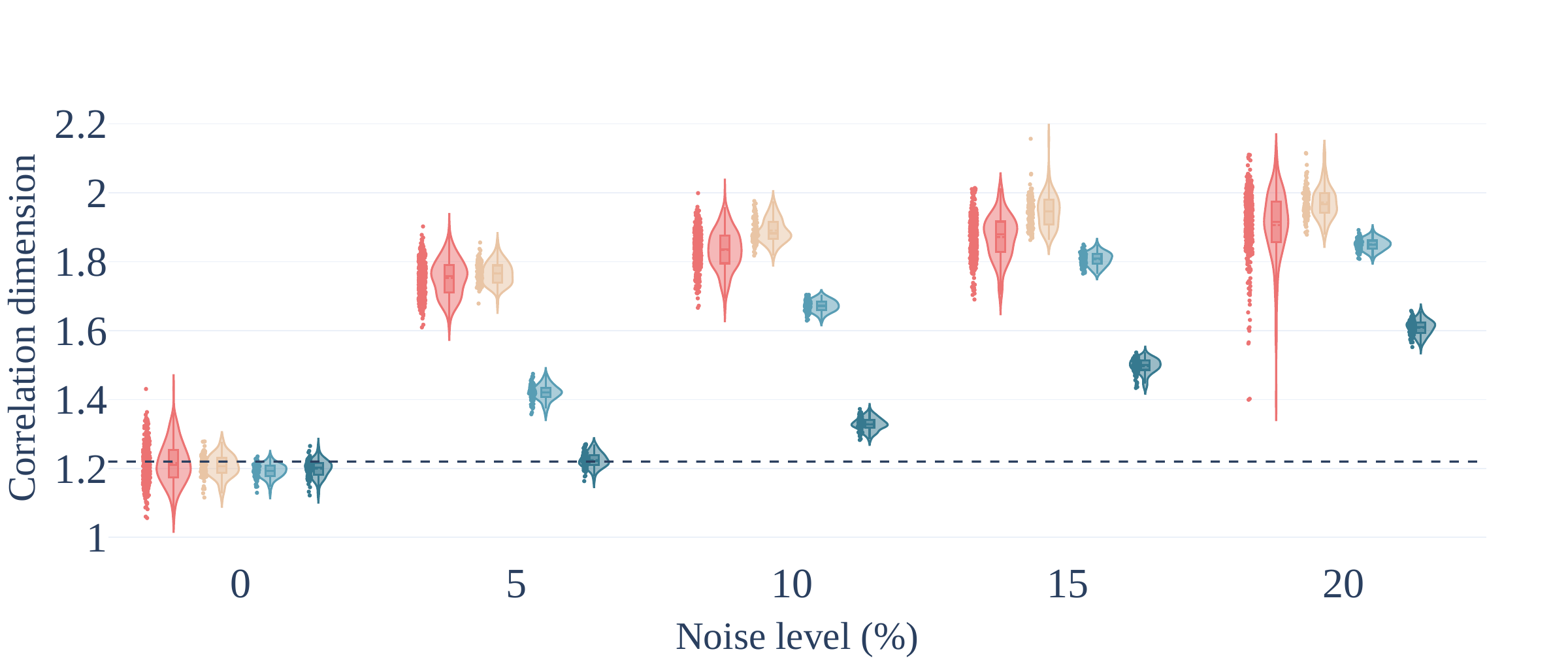}
        \end{subfigure}
        \hfill
        \begin{subfigure}{0.03\textwidth}
            \caption{\label{fig:d2violins_beta_noise_HenonMap}}
        \end{subfigure}
        \caption{Estimation of the correlation dimension from Lorenz (\subref{fig:d2violins_beta_noise_Lorenz}), R\"ossler (\subref{fig:d2violins_beta_noise_Rossler}) and H\'enon (\subref{fig:d2violins_beta_noise_HenonMap}) systems under different levels of additive white Gaussian noise: comparison of the original Grassberger and Proccacia algorithm 
        (blue) with an estimation of the slope in the range $\left[\beta r_\text{opt}, r_\text{opt}\right]$, with $\beta = 0.01$ (beige), $\beta=0.1$ (light blue), $\beta = 0.5$ 
        (dark blue).}  
        \label{fig:corr_dim_violin_noise}
    % \vspace{-0.5em}
    \end{figure}

        We generate 100 time series of 1000 points for the three systems. Each series, 
        with standard deviation $\sigma$, is corrupted with additive white Gaussian noise with standard deviation $\sigma_\text{noise} = k\,\sigma$, where $k$ defines the noise level. 
        As above, we compare the estimation of the original 
        Grassberger and Proccacia algorithm
        with the estimation in the different ranges (the reference radius \eqref{eq:refrule} is 
        computed for each noise-corrupted series). 
        We present in \autoref{fig:corr_dim_violin_noise} a
        violin plot for noise levels $k=0,0.05,0.1,0.15,0.2$. For both R\"ossler and Lorenz attractors, we observe that
        a noise level of 5\% is sufficient to corrupt estimations with the original version of the Grassberger and Proccacia algorithm (red) or the range $\beta = 0.01$ (beige). 
        In contrast, larger values of $\beta$ yield more consistent results under the different noise conditions. 
        Therefore, this observation suggests that in  noise conditions, the correlation dimension can be more robustly estimated from a smaller range of $r$.
          % % \vspace{-0.5em}

\section{Estimation of Kolmogorov-Sinai entropy using recurrence plots}
    In this section, we investigate the behavior of the goodness-of-fit of an estimator of a nonlinear measure with the radius parameter. We also study the reference radius inherent to~\eqref{eq:refrule} under different conditions. We use the reference radius in the construction of recurrence plots used to estimate the Kolmogorov-Sinai (KS) entropy of H\'enon map and apply similar method to real electroencephalographic (EEG) signals. 
    
    \subsection{Recurrence plots and Kolmogorov-Sinai entropy}
        \subsubsection{Recurrence plots}
        Recurrence plots~\cite{eckmann1987recurrence} display phase-space neighbors as a 2D black-and-white image whose $(i, j)$ element is black 
        if trajectory points $x_i$ and $x_j$ are closer than a fixed radius $\varepsilon$.  More formally, 
        from a phase-space trajectory $\{x_i\}, 1 \leq i \leq n$, 
        a recurrence plot $\textsc{rp}(\varepsilon) \in \mathbb{R}^{n\times n}$ is defined as:
        \begin{align}
            \label{eq:recplot}
            (\textsc{rp}(\varepsilon))_{i,j} = \Theta(\varepsilon - \|x_i - x_j\|_p)
        \end{align}
        where $\Theta(\cdot)$ denotes Heaviside step function, $\|\cdot\|_p$ is a norm, usually either $L_1$, $L_2$, or $L_\infty$. The patterns in recurrence plots reflect properties of the underlying dynamical system and can be quantified using the Recurrence Quantification Analysis (RQA) framework, providing a set of  powerful non-parametric visualization and characterization tools for nonlinear time series analysis. The relationship between recurrence plots (and RQA measures) and the correlation sum intuitively follows~\eqref{eq:recplot}~\cite{grendar2013strong}. Indeed, simple mathematical 
        manipulations show that the \emph{recurrence rate}, defined as the average number of recurrent points in a recurrence plot, is equal to the correlation sum \cite{thiel2003analytical}. 
        
        \subsubsection{Estimating Kolmogorov-Sinai entropy from recurrence plots}
    The \emph{Kolmogorov-Sinai} (KS) or \emph{measure-theoretic} entropy \cite{kolmogorov1985new,sinai1959notion} measures the evolution of uncertainty with the iteration of the map of a dynamical system.
    The lower bound $K_2$, often used as the estimate of KS entropy \cite{faure1998new}, is defined as \cite{grassberger1983estimation}:
        \begin{equation}
            \label{eq:ksentropy}
            K_2 = \lim\limits_{r \to 0} \lim\limits_{m \to \infty}\lim\limits_{n \to \infty} \frac{1}{\Delta t} \log \dfrac{C^m(r, n)}{C^{m+1}(r, n)}
        \end{equation}    $C^m(r, n)$ denotes the correlation sum built from a delay-reconstructed trajectory in a $m$-dimensional $L_\infty$ space. 
    While it is possible to approximate the KS entropy directly from correlation sums \cite{pincus1991approximate, richman2000physiological}, we rather consider the method in~\cite{faure1998new}. The latter approximates the KS entropy from the histogram of diagonal lines of length greater than $m$ in a recurrence plot $\textsc{rp}(\varepsilon)$: 
    \begin{equation}
            \label{eq:hist_rp}
            N^{\varepsilon}(m) = \text{card} \{(i,j) : \forall k \in \{0, \dots, m-1\}, |u_{i+k} - u_{j+k}| < \varepsilon\}
    \end{equation}
    A diagonal of size $m$ on a recurrence plot reflects that 
    two trajectories stayed at a distance smaller than a threshold  $\varepsilon$ for $m$ time-steps, or equivalently that two delay-reconstructed vectors in $m$-dimensional space are close under $L_\infty$ norm. Hence, the histogram of diagonal lines, $N^{\varepsilon}(m)$, captures information similar to the correlation sum from delay-coordinates, $C^m(r,n)$; whereas the parameters $\varepsilon$ and $r$ are analogous in the two quantities. The main advantage of the Faure and Korn method is computational: while $C^m(r,n)$ is computed for several values of the embedding dimension $m$, the histogram $N^{\varepsilon}(m)$ is computed only once.
     Then, using the diagonal line histograms to rewrite the KS entropy \eqref{eq:ksentropy} as a function of $r = \varepsilon$ gives: 
    \begin{equation}
    \label{eq:k2diag}
        K_2(r) = \lim\limits_{m \to \infty} \lim\limits_{n \to \infty}  \frac{1}{\Delta t} \log \dfrac{N^{r}(m+1)}{N^{r}(m+2)}
    \end{equation}
    Faure and Korn \cite{faure1998new} suggest to evaluate the average slope of a $\log N^r(m)$ vs $m$ plot for various values 
    of $r$. Then, taking the limit $r \to 0$ is supposed to converge to a constant value equal to the KS entropy, $K_2$, up to a scaling factor. However, selecting the smallest possible $r$ to estimate the limit $r\to 0$ from real-world
    data (i.e.\ finite-size samples with noise) likely leads to estimations flawed by a large variance, as discussed in~\cite{faure2015estimating}. 
    Hence, the problem is to select a value of $r$ yielding the best possible estimations of the KS entropy.
    We use our reference radius (\eqref{eq:refrule}) to compute the recurrence plot used to estimate the KS entropy. Recurrence plots and diagonal line histograms were computed using the \texttt{pyunicorn} package \cite{donges2015unified}.
    
    \subsection{Numerical experiments for the H\'enon map}
    
\begin{figure}[t!]
    % \vspace{-0.5em}
        \centering
        \includegraphics[width=0.9\linewidth, trim={0 0 0 4em}, clip=true]{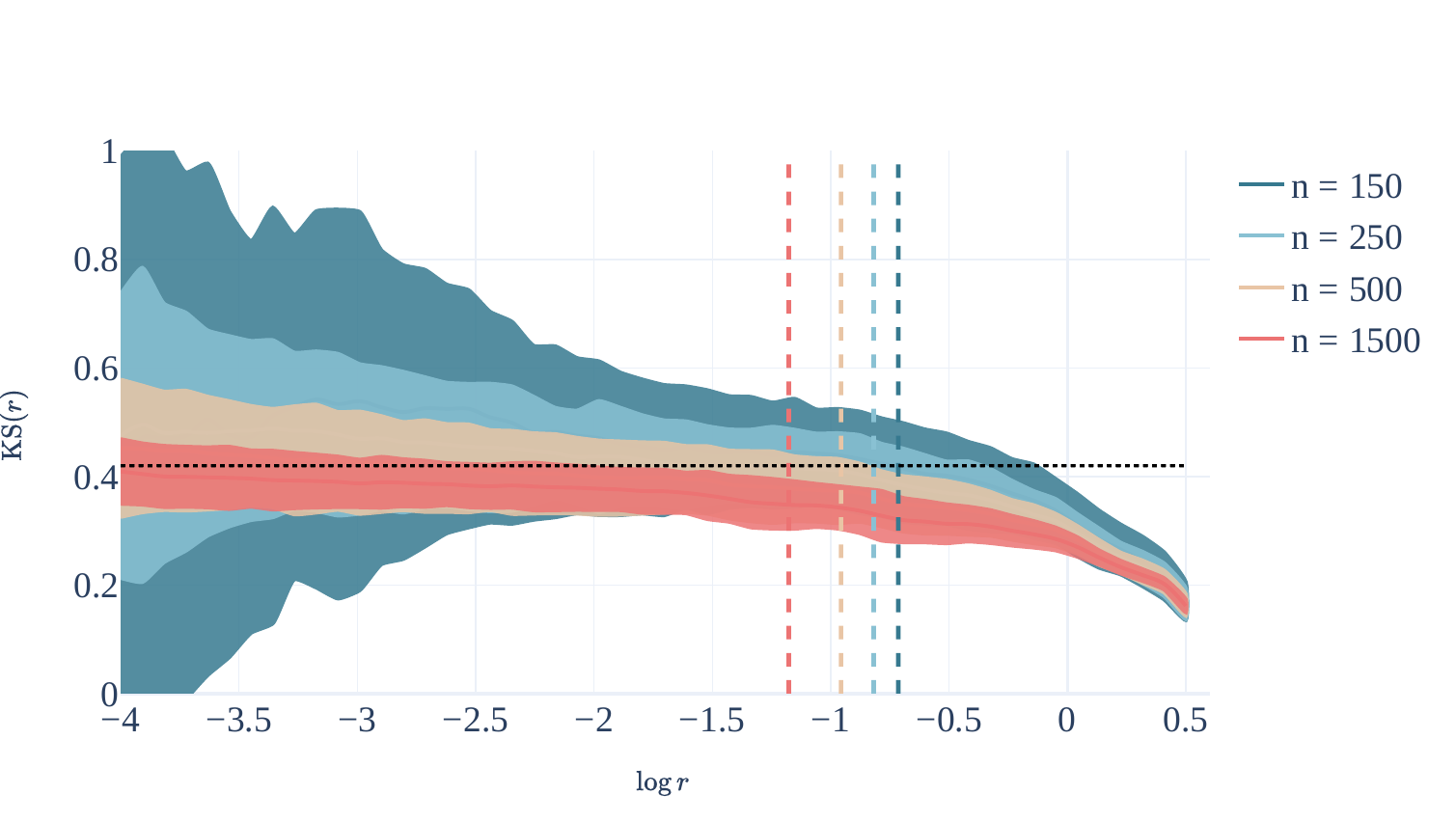}
        \caption{Estimation of Kolmogorov-Sinai entropy for time series from the H\'enon map with different lengths $n$. The filled areas corresponds to the $95\%$ 
        (Gaussian) confidence intervals for each length. The vertical dashed lines represents the average reference radius associated to each length. 
        The horizontal dashed line indicates the reported entropy for H\'enon map, $K_2 = 0.42$.}
        \label{fig:ks_entropy}
    % \vspace{-0.5em}
    \end{figure}

    We generate 100 time series for each length ($n = 150, 250, 500, 1500$ points) from the standard H\'enon map ($a = 1.4$, $b = 0.3$). For each series, we use the 
    Faure and Korn method and compute the value of $K_2(r)$, \eqref{eq:k2diag}, as a function $\log r$ curve for $50$ values of $\log r$ ranging from $-4$ to $0.5$. 
    We then compute the reference radius (\eqref{eq:refrule}) --- using the series length $n$ and dimension $d=1$ --- and average the values over series of same length.
    Results are presented in \autoref{fig:ks_entropy}.
    
        We notice that the variance of the estimation increases for 
        decreasing radius and 
        decreasing number of points. This result is presumably due to a poor statistical power for small values of the radius and short time series. 
        However, the estimation seems to converge
        in average to the theoretical value ($H_{KS} = 0.42$~\cite{faure1998new}) when $r$ tends to $0$.
        Notice that the right-most part of the plot exhibits a 
        large bias between the estimated and theoretical entropy values. Contrary to 
        the variance, the bias does not seem to decrease with increasing number of points. Thus this bias is more symptomatic of the radius being too high to obtain any valuable information about the H\'enon map. This bias-variance trade-off is usually related to a Mean Squared Error (MSE) minimization problem.  
       
    \begin{figure}[!htb]
    % \vspace{-0.5em}
        \centering
        \includegraphics[width=0.9\linewidth, trim={0 0 0 3em}, clip=true]{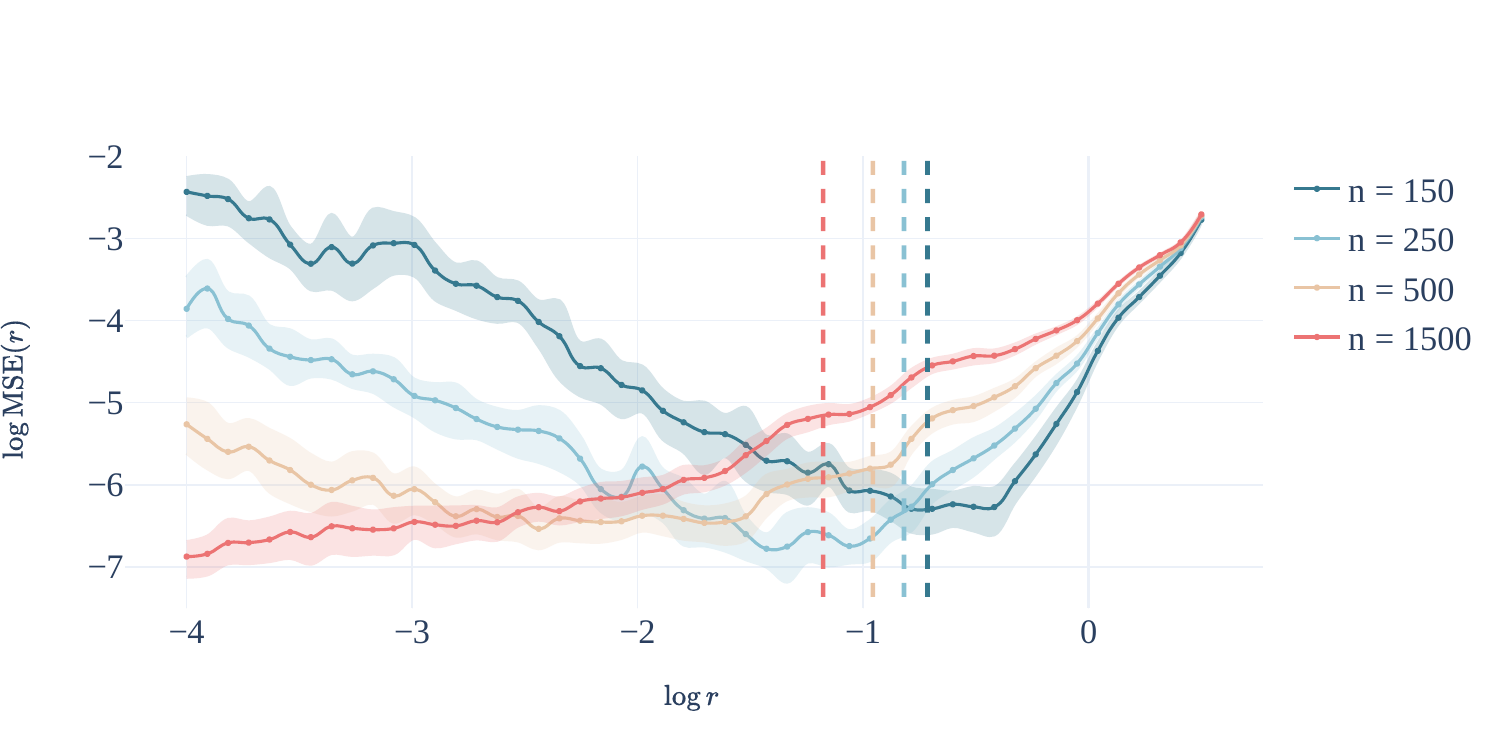}
        \caption{Estimation of the log-Mean Square Error of the Kolmogorov-Sinai entropy estimator as a function of the radius value (lower is better). 
        The filled areas corresponds to the $95\%$ bootstrap confidence interval for each length.
        We see that for different time series, the reference radius gives a log-MSE value between $-7$ and $-5$.}
        \label{fig:ks_entropy_mse}
    % \vspace{-0.5em}
    \end{figure}
    
        As the MSE of an estimator quantifies the goodness-of-fit, the parameters of the estimator yielding the minimum value of MSE can be systematically selected. 
        We numerically compute the MSE of the KS entropy estimator as a function of $\log r$ and use this plot as an objective criterion to evaluate the adequacy of our reference radius. The MSE consists in the sum of a squared bias term,
        measuring the difference between the theoretical value and the estimation, as well as the variance of the estimator. We use a theoretical value $K_2 = 0.42$ and all of the 100 sample series to compute 
        the MSE, overlay the reference radius averaged over series of same length, and show the results in \autoref{fig:ks_entropy_mse}.
        For short time series, we observe that the radius selected by the reference rule is
        systematically close to the minimum of the MSE. For longer time series, the reference radius is larger than the minimum
        of the curve. Nevertheless, for values of $r < r_\text{opt}$, the slope of the MSE curve gets flatter for increasing number of points and allows arbitrary selection of smaller radius
        values.
        We report similar observations for two other estimators of the KS entropy, the Approximate and Sample entropies (results not shown).
        
    \subsection{Application to EEG signals in the context of epilepsy}
        To show the viability of our approach on real-world data, we apply our radius selection procedure to estimate the KS entropy of epileptic EEG signals. 
        A significant decrease of the EEG signal entropy at the epileptic seizure location is a common feature for automatic seizure detection~\cite{ocak2008optimal, srinivasan2007approximate}. 
        We use the data publicly available from the University of Bonn~\cite{andrzejak2001indications}, which consists in five sets of EEG data. Each set contains 100 segments of $23.6$ seconds recorded at $173.61$Hz (4096 points per segment), which were 
        visually inspected for artifacts and band-pass filtered between $0.5$Hz and $40$Hz. 
        Two sets contains surface EEG recorded from five healthy volunteers at rest, either with closed (set O) and opened eyes (set Z). The three other sets, consisting in signals from five epileptic patients recorded during presurgical evaluation, 
        contain segments either from seizure-free intervals (at epileptogenic site, set F, or at the hippocampal formation of the opposite hemisphere of the brain, set N) or during seizure (at epileptogenic site, set S).
       
    \begin{figure}[ht!]
    \vspace{-0.5em}
            \begin{subfigure}[b]{0.4\linewidth}
                \includegraphics[ width=\linewidth, trim={0 0 0 2em}, clip=true]{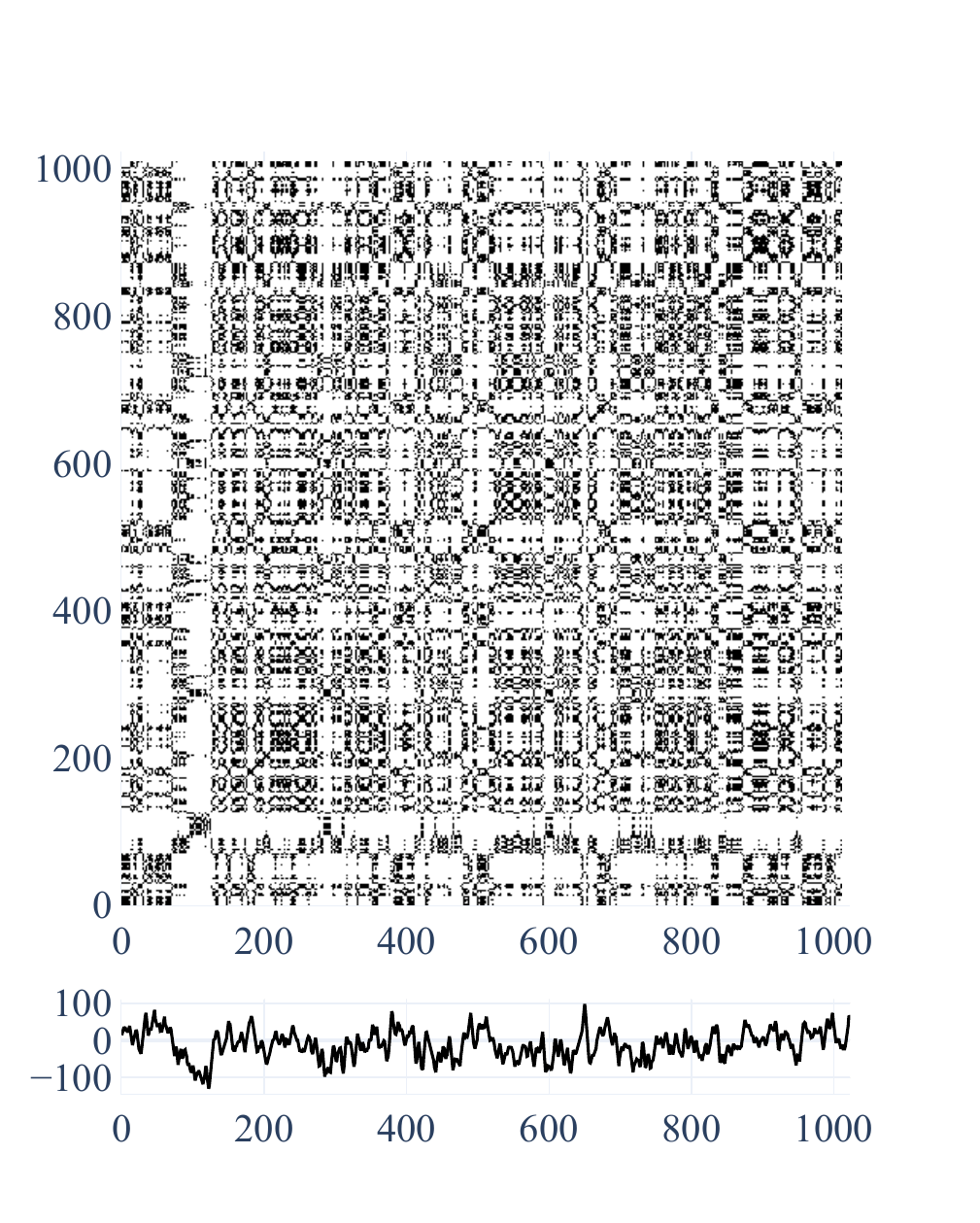}
                \vspace{-3em}
                    \caption{\label{fig:rp_eeg_bl_z}}
                \end{subfigure}% \hfill
            \begin{subfigure}[b]{0.4\linewidth}
                \includegraphics[ width=\linewidth, trim={0 0 0 2em}, clip]{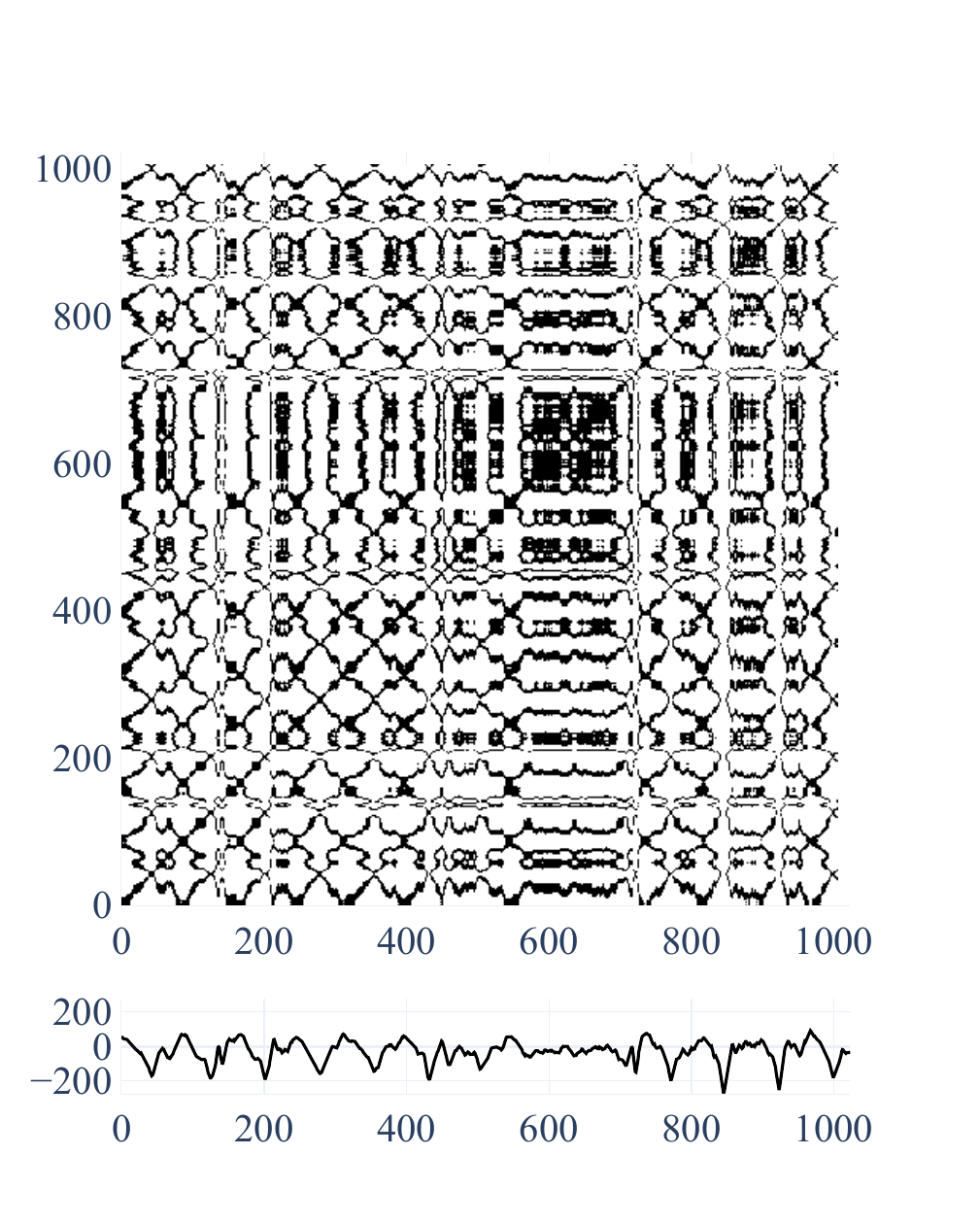}
                \vspace{-3em}
                \caption{\label{fig:rp_eeg_epi_f}}
            \end{subfigure}
            \hfill
        \begin{subfigure}[b]{0.19\textwidth}
                \centering
                \includegraphics[width=\linewidth, trim={0 0 0 2em}, clip]{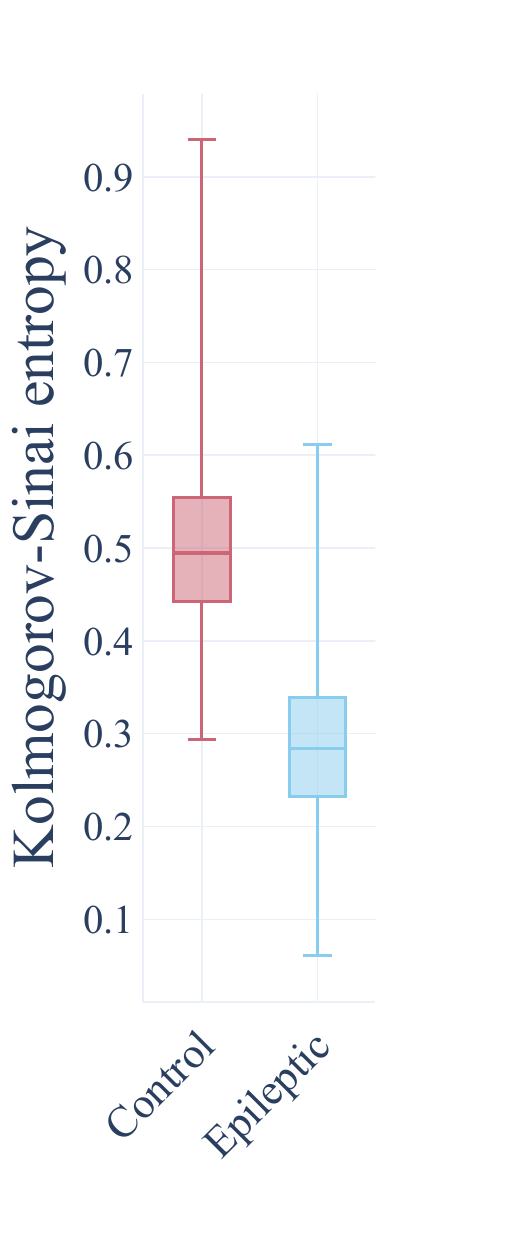}
                \vspace{-3em}
                \caption{\label{fig:ks_eeg_epilepsy}}
            \end{subfigure}
            \caption{Estimation of the Kolmogorov-Sinai entropy from recurrence plots to discriminate epileptic from healthy EEG signals: (\subref{fig:rp_eeg_bl_z}) and (\subref{fig:rp_eeg_epi_f}) show sample EEG signals and recurrence plots for an healthy volunteer and an epileptic patient respectively. The value $r = 1.843 \times \hat{s} \times n^{-1/5}$ (\eqref{eq:refrule}) is used to compute the recurrence plots from the univariate time series; (\subref{fig:ks_eeg_epilepsy}) contains a box plot of the estimated entropies for both control and epileptic groups. As expected, the entropy values are significantly lower for the epileptic group.}
    \vspace{-0.5em}
        \end{figure}

    Each record is divided in four segments of 1024 points. For each segment, we compute a recurrence plot with the radius set by \eqref{eq:refrule} and estimate the KS entropy using the Faure and Korn method.
    Recurrence plots and signals sampled from the sets Z and F are shown in \autoref{fig:rp_eeg_bl_z} and \autoref{fig:rp_eeg_epi_f}. We present in \autoref{fig:ks_eeg_epilepsy} a box plot of the KS entropy for the healthy volunteers (control group) and the epileptic patients. 
    Our estimator gives an average KS entropy of $0.288\pm0.005$ (95\% confidence interval) for the epileptic group and $0.504\pm0.006$ for the control group, which confirms an average significant decrease of the KS entropy with epilepsy, as reported in previous studies \cite{kannathal2005entropies}. 
    
    Finally, to compare the discrimination strength of common closed-form radius selection methods, we estimate the KS entropy with each method, perform a two-samples Z-test (epileptic versus control group) and collect the Z-score. 
    We report a Z-score of $Z = 45.3$ (resp.\ $Z = 39.3$) for the $r = 0.2\sigma$ (resp.\ $r = 0.1\sigma$, with $\sigma$ the series standard deviation) 
    rule \cite{pincus1991approximate}, $Z = 48.9$ when the radius is set to $10\%$ of the maximum phase space neighborhood \cite{zbilut1992embeddings}, $Z = 41.1$ (resp. $Z = 34.6$) when the radius is selected 
    such that $10\%$ (resp.\ $4$\%) of the number of points are selected as neighbors \cite{webber2015recurrence,kraemer2018recurrence}, 
    $Z = 54.9$ for the reference rule radius $r=1.843 \times \hat{s} \times n^{-1/5}$ (\eqref{eq:refrule}). Subsequently, although all methods detect significant differences between the two groups, the radius given by~\eqref{eq:refrule} gives the most statistically significant results. 
    
\section{Discussion and conclusion}
    We propose a new approach for selecting the radius parameter in nonlinear measures derived from the correlation sum. We first formulate a relative error function on the quantities underlying correlation sums. We show that minimizing the loss function is equivalent to minimizing the MISE of a kernel density estimator. We use the AMISE minimization method to derive a closed-form expression 
    to select the radius. Additionally, we observe how the bias and variance of the estimator varies with 
    the radius and derive a ``meaningful'' range to select a variable radius. 
      
    We investigate the behavior of the Grassberger and Proccacia algorithm for estimating the correlation dimension in radius ranges of different size.
    We observe that the range parameter $\beta$ can be selected close to $1$ for low-variance estimations, and close to $0$ for low-bias
    estimations. However, the presence of noise in the observed signal induces typical error in the estimations and leads to favor 
    small ranges close to the reference radius. 
    
    We then use the reference radius to construct recurrence plots for estimating the Kolmogorov-Sinai entropy from both simulated and experimental signals. 
    In a first analysis, we reconstruct the Mean Squared Error curve of the entropy estimator for H\'enon map and show that the reference radius is close to the minimum of the curve. 
    We confirm the experimental adequacy of the method by obtaining significant results in characterizing epileptic EEG signals. 
    
    Moreover, our theoretical approach yields a reference radius that is similar to several existing radius selection methods arising from 
    empirical or numerical experience: the radius is a fraction of the scale of the data \cite{pincus1991approximate,zbilut1992embeddings}
    and compensates for the dimension of the data \cite{kraemer2018recurrence}.
    
    For the specific case of recurrence plots, \cite{andreadis2020topological}
    recently proposed an empirical procedure to identify an optimal radius value. They define a metric to measure the distance between recurrence plots and compute the distance between recurrence plots constructed from the same time series using increasing values of radius. The radius value is considered ``optimal'' when it minimizes the distance between consecutive recurrence plots, i.e. such that a slightly changing the radius has the minimal impact on the recurrence plot. The principal issues with this procedure are the computational burden of building several recurrence plots and the difficulty to reliably identify the optimum. In contrast, our method is computationally
    much more efficient and not restricted to recurrence plots.  
    % theoretical justification, no asumption on the dynsys (stoch vs determ).   
    
    % Although the method is adequate for practical applications where the analyzed signals are short and/or noisy, the reference radius might be overly large for long time series with low levels of observational noise,
    % or for systems exhibiting different dynamical properties at different amplitudes (see \ref{section:corrdim_results_range}).  
    
     Our numerical experiments suggest that the reference radius given in~\eqref{eq:refrule} can be used as a default parameter to obtain robust and significant values for a number of different nonlinear tools and measures: correlation dimension, recurrence plots, Kolmogorov-Sinai entropy. In future work, we plan to investigate the relation between our optimal radius and the embedding parameters, which play a role on the trajectories resolution in the reconstructed phase space. Additionally, we plan to use the reference radius in EEG signal processing application, notably to extract dynamical features characterizing the oscillatory dynamics of motor imagery EEG signals.  
    
\section*{Acknowledgements} 
    This research was supported by a grant from the French Minist\`ere de l’\'Education Nationale, de l’Enseignement Sup\'erieur et de la Recherche. The code related to this work is made openly available under an MIT license at \href{https://github.com/johmedr/pykeos}{https://github.com/johmedr/pykeos}.    

% \section*{Data availability}
%     Data sharing is not applicable to this article as no new data 
%     were created or analyzed in this study.
    
\bibliographystyle{rss}
\bibliography{references}

\appendix

% \section{Mathematical details}
\section{Volume of generalized balls}
\label{section:app_ball}
Let $B_r^{p,d}(x)$ be an open $d$-ball of size $r$ in an $L_p$ space, i.e. 
$B_r^{p,d}(x) = \{y \in \mathbb{R}^d : \|x - y \|_p < r\}$. We write $\tau_{p, d} = \lambda(B_1^{p,d})$ the volume of the generalized unit ball, where 
$\lambda$ is the Lebesgue measure. The volume of a generalized ball of radius $r$ is $\lambda(B_r^{p,d}) = \tau_{p, d} r^d$.
The general formula for the volume of a generalized unit ball is \cite{wang2005volumes}: 
\begin{equation}
    \tau_{p, d} = \frac{(2 \Gamma(\frac{1}{p} + 1))^d}{\Gamma(\frac{d}{p} + 1)}
\end{equation}
which can be simplified for common $L_p$ spaces: 
\begin{align}
    \tau_{1, d} &= \frac{2^d}{d!} & \tau_{2, d} &= \frac{\pi^{\frac{d}{2}}}{\Gamma(\frac{d}{2} + 1)} &  
    \tau_{\infty, d} &= 2^d
\end{align}
$\Gamma$ denotes Euler's gamma function with the property $\Gamma(z+1) = z\Gamma(z)$.

% \section{Uniform kernel in $d$-dimensional $L_p$ space}
% \label{section:app_kernel}
%  Notice that $\int_{\mathbb{R}^d} K_h(u) \mathrm{d}u =1$. 

\section{Derivation of a reference rule for the uniform kernel}
\label{section:app_refrule}
The expression of the bandwidth minimizing the Asymptotic Mean Integrated Squared Error 
is (\cite{silverman1986density}, Eq. 4.14 and 4.15):
\begin{equation}
                \label{eq:app_h_AMISE}
                h_{\text{AMISE}} = \left[\dfrac{W_1(K) \cdot d}{n  \cdot \left[W_2(K)\right]^{2} \cdot  W_1(\nabla^2f)}\right]^{1/(d+4)} 
            \end{equation} 
where $W_i$ are the functionals $W_1(g) = \int_{\mathbb{R}^d} g^2(u) du$ and
$W_2(g) = \int_{\mathbb{R}^d} u_1^2 g(u) du$, where $u_1$ is the first component of $u \in \mathbb{R}^d$ (as the kernel is symmetric, it is sufficient 
to consider only $u_1$ in $W_2$).
We compute $W_1$ for the uniform kernel: 
\begin{align}
    W_1(K) = \int\limits_{\mathbb{R}^d} K^2(u) du = \int\limits_{B_1^{p,d}(0)} \left(\frac{1}{\tau_{p, d}}\right)^2 du  = \frac{1}{\tau_{p, d}}
\end{align}
$W_1(\nabla^2\rho)$ for a $d$-dimensional Gaussian reference distribution $\phi$ is given in~\cite[Eq. 4.13]{silverman1986density}: 
\begin{align}
     W_1(\nabla^2\rho) \approx W_1(\nabla^2\phi) = (2\sqrt{\pi})^{-d} \left(d/2 + d^2/4\right)
\end{align}
   
Then, $W_2$ in the 1-dimensional case: 
\begin{align}
    W_2(K) = \int\limits_{\mathbb{R}} u^2 K(u) du = \int\limits_{-1}^{1} \frac{u^2}{\tau_{p, 1}} du = \frac{1}{3}
\end{align}
For $d \geq 2$, using $u_i$ to denote the $i$-th coordinate of $u \in \mathbb{R}^d$:
\begin{align*}
    W_2(K) & = \int\limits_{\mathbb{R}^d} u_1^2 K(u) du \\
    &= \frac{1}{\tau_{p, d}} \int\limits_{\mathbb{R}}u_1^2 \left(\,\, \int\limits_{\mathbb{R}^{d-1}} 
    \Theta(1 - (|u_1|^p + \sum_{i=2}^d |u_i|^p)^{1/p}) du_d \dots du_2\right) \, du_1 
\end{align*}
Changing to spherical coordinates $\vec{u}_{2:d} = (\eta, \vec{\xi})$ with an orientation vector $\vec{\xi}$ and a radius $\eta =  (\sum_{i=2}^d |u_i|^p)^{1/p}$ and :
\begingroup
\allowdisplaybreaks
\begin{align*} 
      W_2(K)  &= \frac{1}{\tau_{p, d}} \int\limits_{\mathbb{R}} u_1^2
    \int\limits_{\eta = 0}^\infty   \int\limits_{\mathbb{R}^{d-2}}
    \Theta\left(1 - (|u_1|^p + \eta^p)^{1/p}\right) d\vec{\xi} \, d\eta \,  du_1  \\
    &= \frac{\tau_{p, d-1}}{\tau_{p, d}} \int\limits_{\mathbb{R}} u_1^2
    \int\limits_{\eta=0}^\infty  (d-1)  \eta^{d-2}  
    \Theta\left(1 - (|u_1|^p + \eta^p)^{1/p}\right) d\eta\, du_1  \\
    &= \frac{\tau_{p, d-1}}{\tau_{p, d}} \int\limits_{u_1=-1}^1 u_1^2
    \int\limits_{\eta=0}^{(1 - |u_1|^p)^{1/p}}
     {(d-1)}  \eta^{d-2} d\eta\, du_1  \\
     &= \frac{\tau_{p, d-1}}{\tau_{p, d}} \int\limits_{u_1=-1}^1 u_1^2 \left((1 - |u_1|^p)^{1/p}\right)^{(d-1)}
     \,du_1 \\
     &= \frac{\tau_{p, d-1}}{\tau_{p, d}} \frac{2}{3} {}_2F_1\left(\frac{3}{p}, 
    \frac{1-d}{p};\frac{3+ p}{p};1\right) 
\end{align*} 
\endgroup
where ${}_2F_1(a,b;c;z)$ is the Gaussian hypergeometric function.
% \footnotemark.  
% \footnotetext{This result was obtained using Wolfram Alpha with the query ``int x\textasciicircum2 (1-x\textasciicircum p)\textasciicircum((d-1)/p)'' and checked manually (Wolfram Alpha LLC, 2018. Wolfram|Alpha., accessed on 01/26/2021).}
Finally, using the fact that $\frac{\tau_{p, d-1}}{\tau_{p, d}} = \frac{\Gamma(\frac{d}{p} + 1)}{2 \Gamma(\frac{1}{p} + 1) \Gamma(\frac{d-1}{p} + 1)}$, (see Appendix~A.1) and the expansion of ${}_2F_1$ at $z = 1$ \cite[Eq.~15.4.20]{DLMF}, we can further simplify: 
\begin{align}
    W_2(K) &= \frac{\tau_{p, d-1}}{\tau_{p, d}} \frac{2 \Gamma\left(1 + \frac{3}{p}\right) \Gamma\left(\frac{d-1}{p}+1\right)}{3 \Gamma\left(\frac{d+2}{p} + 1\right)}
    = \frac{\Gamma\left(\frac{d}{p} + 1\right) \Gamma\left(1 + \frac{3}{p}\right)}{3 \Gamma\left(\frac{d+2}{p} + 1\right) \Gamma\left(\frac{1}{p} + 1\right)}
\end{align}

We then derive the reference rule by plugging the appropriate values in \eqref{eq:app_h_AMISE}. 
First, we derive the expression in the simple 1 dimensional case, which is independent from $p$: 
\begin{align}
    r_{\text{opt}} = 
    \left[ 12\sqrt{\pi} \right]^{1/5} \cdot \hat{s} \cdot n^{-1/5}  \approx 1.843 \cdot \hat{s} \cdot n^{-1/5} 
\end{align}
The general formula for $d \geq 2$ is more complex:
\begin{align}
    r_{\text{opt}}&= \left[\dfrac{4 d (2\sqrt{\pi})^2}{\tau_{p, d} d (d+2)} \times \dfrac{\left(3 \Gamma\left(\frac{d+2}{p} + 1\right) \Gamma\left(\frac{1}{p} + 1\right) 
    \right)^2}{\left(\Gamma\left(\frac{d}{p} + 1\right) \Gamma\left(1 + \frac{3}{p}\right)\right)^2}\right]^{1/(d+4)}  
     \cdot \hat{s} \cdot n^{-1/(d+4)} \nonumber \\
     &= \left[ \dfrac{4 (2\sqrt{\pi})^{d} \left(3 \Gamma\left(\frac{d+2}{p} + 1\right) \Gamma\left(\frac{1}{p} + 1\right)
    \right)^2}{\tau_{p, d} (d+2)\left(\Gamma\left(\frac{d}{p} + 1\right) \Gamma\left(1 + \frac{3}{p}\right)\right)^2}\right]^{1/(d+4)} 
     \cdot \hat{s} \cdot n^{-1/(d+4)} \label{eq:app_genexpr}
\end{align}

\section{Simplification of the reference rule for common norms}
\label{section:app_splfy}
We address the case $p = 1$: 
\begin{align}
    r_{\text{opt}}&= \left[ \dfrac{4 (2\sqrt{\pi})^{d} \left(3 \Gamma\left(d + 3\right) \Gamma\left(2\right)
    \right)^2}{\tau_{p, d} (d+2)\left(\Gamma\left(d + 1\right) \Gamma\left(4\right)\right)^2}\right]^{1/(d+4)} 
     \cdot \hat{s} \cdot n^{-1/(d+4)} \nonumber\\ 
      &= \left((\sqrt{\pi})^d (d+1) \left(d+2\right) ! \right)^{1/(d+4)}
     \cdot \hat{s} \cdot n^{-1/(d+4)} \label{eq:app_l1expr}
\end{align}
Then, the limiting case $p\to\infty$: 
\begin{align}
    r_{\text{opt}}&= \left[ \dfrac{4 (2\sqrt{\pi})^{d} \left(3 \Gamma\left(1\right) \Gamma\left(1\right)
    \right)^2}{\tau_{p, d} (d+2)\left(\Gamma\left(1\right) \Gamma\left(1\right)\right)^2}\right]^{1/(d+4)} 
     \cdot \hat{s} \cdot n^{-1/(d+4)} \nonumber \\ 
     &= \left[ \frac{36 (\sqrt{\pi})^d }{d+2} \right]^{1/(d+4)}
     \cdot \hat{s} \cdot n^{-1/(d+4)}  \label{eq:app_l2expr}
\end{align}
Finally the case $p = 2$: 
\begin{align}
    r_{\text{opt}}&= \left[ \dfrac{4 (2\sqrt{\pi})^{d} \left(3 \Gamma\left(\frac{d+2}{2} + 1\right) \Gamma\left(\frac{1}{2} + 1\right)
    \right)^2}{\tau_{2, d} (d+2)\left(\Gamma\left(\frac{d}{2} + 1\right) \Gamma\left(1 + \frac{3}{2}\right)\right)^2}\right]^{1/(d+4)} 
     \cdot \hat{s} \cdot n^{-1/(d+4)} \nonumber\\
     &= \left[ \dfrac{2^{d+2} \Gamma\left(\frac{d}{2} + 1\right)  \left(3 \Gamma\left(\frac{d}{2} + 1\right) (\frac{d}{2} + 1) \Gamma\left(\frac{1}{2} + 1\right)
    \right)^2}{(d+2)\left(\Gamma\left(\frac{d}{2} + 1\right) \Gamma\left(1 + \frac{1}{2}\right) (1 + \frac{1}{2})\right)^2}\right]^{1/(d+4)} 
     \cdot \hat{s} \cdot n^{-1/(d+4)} \nonumber \\
     &= \left[2^{d+2} \Gamma\left(\frac{d}{2} + 1\right) (d+2)\right]^{1/(d+4)} 
     \cdot \hat{s} \cdot n^{-1/(d+4)}\nonumber \\
      &= 2 \left[\frac{\Gamma\left(\frac{d}{2} + 1\right)}{2}\right]^{1/(d+4)} 
     \cdot \hat{s} \cdot n^{-1/(d+4)}  \label{eq:app_linfexpr}
\end{align}

\end{document}